\documentclass[british]{scrreprt}
\usepackage[utf8]{inputenc}
\usepackage[T1]{fontenc}
\usepackage{libertine}

\usepackage{babel}
\usepackage{amsfonts, amsmath, amssymb, amsthm, bbm, color}
\usepackage{mathrsfs}           %script capital math (\mathscr)
\usepackage{xcolor}
\usepackage{u8chars}
\usepackage{relsize}
\usepackage{xspace}
\usepackage[normalem]{ulem}     %for \sout…
\usepackage[group-separator={,},group-four-digits]{siunitx}
\usepackage{etoolbox}
\usepackage{comment}            %exclude/include/special-comment
\usepackage[inline,shortlabels]{enumitem}
\usepackage{url}
\usepackage[numbers]{natbib}
\usepackage[textsize=small]{todonotes}
\usepackage{tikz}
\usepackage{graphicx}
\usepackage[frozencache,cachedir=.]{minted}

\usepackage{scrhack}            %suppress warning about \float@addtolists
\usepackage{tcolorbox}
\usepackage{changepage}
\usepackage[%
colorlinks,
linkcolor={red!50!black},
citecolor={blue!50!black},
urlcolor={blue!80!black},
]{hyperref}
\usepackage{microtype}
\DisableLigatures{encoding = T1, family = tt*}

\DeclareMathOperator*{\argmax}{arg\,max}

\usepackage{xifthen}% provides \isempty test
\newcommand\lambdaone[1]{\ensuremath{\ifthenelse{\isempty{#1}}{}{（#1）}}}
\newcommand\lambdatwo[2]{\ensuremath{\ifthenelse{\isempty{#1#2}}{}{（#1, #2）}}}

\newcommand\featcomp[2]{\ensuremath{\FF_{{#1},{#2}}}\xspace}

\newcommand{\red}[1]{{\leavevmode\color{red}#1}\xspace}%

\newcommand{\cf}		{{\itshape	      cf.}\xspace}

\newcommand{\eg}		{{\itshape	    e.g.,}\xspace}

\newcommand{\ie}		{{\itshape	    i.e.,}\xspace}

\newcommand{\wrt}		{{\itshape	   w.r.t.}\xspace}
\newcommand{\st}		{{\itshape	    s.t\/}\xspace}

\newcommand{\DeepConcolic}{\textsf{DeepConcolic}\xspace}
\newcommand{\TestRNN}{\textsf{TestRNN}\xspace}
\newcommand{\EKiML}{\textsf{EKiML}\xspace}
\newcommand{\GUAP}{\textsf{GUAP}\xspace}

\newcommand\testrnn{\TestRNN}

\newcommand\dbnabstr{\texttt{dbnabstr}\xspace}

\newcommand\B{\ensuremath{\mathcal B}\xspace}%
\newcommand\N{\ensuremath{\mathcal N}\xspace}%
\newcommand\FF{\ensuremath{\mathbb F}\xspace}%
\newcommand\RR{\ensuremath{\mathbb R}\xspace}%

\newcommand\layer[1]{\ensuremath{l_{#1}}\xspace}

\newcommand\subfeatspace[3]{\ensuremath{f^{♯#3}_{#1,#2}}\xspace}
\newcommand\subfeatlb[3]{\ensuremath{f^{[#3}_{#1,#2}}\xspace}

\newcommand\LF[1][h]{#1idden feature\xspace}
\newcommand\LFs[1][h]{#1idden features\xspace}
\newcommand\LFI[1][h]{#1idden feature interval\xspace}
\newcommand\LFIs[1][h]{#1idden feature intervals\xspace}
\newcommand\HF[1][h]{#1idden feature\xspace}
\newcommand\HFs[1][h]{#1idden features\xspace}
\newcommand\HFI[1][h]{#1idden feature interval\xspace}
\newcommand\HFIs[1][h]{#1idden feature intervals\xspace}

\newcommand\BNa[2]{\ensuremath{\B_{#1,#2}}\xspace}%
\let\BN=\BNa
\newcommand\LInf[2]{\ensuremath{\|#1 - #2\|_{∞}}\xspace}%

\newcommand\Xtrain{\ensuremath{X_{\mathit{train}}}\xspace}%
\newcommand\Ytrain{\ensuremath{Y_{\mathit{train}}}\xspace}%
\newcommand\Xtest{\ensuremath{X}\xspace}%
\newcommand\Ytest{\ensuremath{Y}\xspace}%

\newcommand\dthr{\ensuremath{d_{\mathrm{thr}}}\xspace}%
\newcommand\dminHard{\ensuremath{d_{\min}}\xspace}%
\newcommand\dminNoise{\ensuremath{d_{\min}^+}\xspace}%

\newcommand\Nhs{\lstinline[language=sh,basicstyle=\ttfamily\small]%
  {har_dense.h5}\xspace}%
\setlist{noitemsep}
\setlist[1]{labelindent=\parindent} % < Usually a good idea
\newlist{compactitem}{itemize}{4}
\setlist[compactitem,1]{nolistsep,label=\textbullet}
\setlist[description]{font=\mdseries\itshape}
\newlist{mathdesc}{description}{4}
\newlist{mathpardesc}{description}{4}
\newlist{mathpardesc*}{description}{4}
\newlist{mathpardesc**}{description*}{4}
\newcommand*{\keymathbox}[1]{%
  \mdseries%
  \upshape%
  \setlength{\fboxsep}{.4pt}%
  \fcolorbox{gray}{white}{%
    \strut\ensuremath{#1}%
  }%
}%
\setlist[mathdesc]{format=\ensuremath}
\setlist[mathpardesc]{format=\keymathbox,% , labelindent=1em
  leftmargin=\parindent,
  labelindent=0pt,
}
\setlist[mathpardesc*]{format=\keymathbox,
  nosep,
  leftmargin=0pt,
  labelindent=\parindent,
}
\setlist[mathpardesc**]{format=\keymathbox,mode=unboxed}

\newlist{bolddescr}{description}{4}
\newcommand*{\mybolddescritem}[1]{\bfseries\upshape{#1}:}%
\setlist[bolddescr]{style=sameline, nosep, format=\mybolddescritem,
  leftmargin=0pt, labelindent=0pt,
}

\newcommand*\maindescfmt[1]{\upshape\bfseries #1:}%
\newlist{maindesc}{description}{4}
\setlist[maindesc]{%
  nosep,
  format=\maindescfmt,
  style=nextline,
  labelindent=0pt,
  leftmargin=\parindent,
}

\newlist{subdesc}{description}{4}
\setlist[subdesc]{%
  nosep,
  style=sameline,
  format=\mdseries\itshape\(\circ\)~,
  leftmargin=\parindent,
  labelindent=2pt,
}

\usetikzlibrary{positioning, intersections, fit, calc}
\usetikzlibrary{arrows, arrows.meta, backgrounds}
\tikzset{
  every path/.style = {
    line width=1pt,
    cap=round,
    join=round,
  },
}
\tcbuselibrary{skins}
\tcbuselibrary{breakable}

\tcbset{%
  textmarker/.style={%almost straight from tcolorbox's manual
    breakable,
    oversize,
    parbox=false,
    boxrule = 0mm,
    leftrule = 1mm,
    rightrule = 1mm,
    boxsep = 0mm,
    arc = 0mm,
    outer arc = 0mm,
    left = 2mm,
    right = 2mm,
    top = 1mm,
    bottom = 1mm,
  }%
}

\newtcolorbox{warn}{%
  textmarker,
  colback = yellow!5!white,
  colframe = yellow%
}

\newtcolorbox{logbox}{%
  textmarker,
  colback = shback!66!white,
  colframe = shframe!66!white,
  leftrule = 1pt,
  rightrule = 1pt,
}

\tcbuselibrary{listings}
\tcbuselibrary{skins}

\colorlet{shframe}{black!33!white}
\colorlet{shback}{black!6!white}

\newcommand\lstshellstyle{% \fontfamily{lmtt}\selectfont%
  \ttfamily
  \upshape%
  \footnotesize% \ttfamily
}

\usepackage{accsupp}
\newcommand\NotCopyable[1]{\BeginAccSupp{method=escape,ActualText={}}#1\EndAccSupp{}}

\newtcblisting{cmds}{%
  textmarker,
  colframe = shframe,
  colback = shback,
  breakable = false,            %<- for copy-paste
  listing only,%
  listing options={%
    style=tcblatex,
    language=sh,
    prebreak=\lstshellstyle\textbackslash,
    basicstyle=\lstshellstyle,
  },
  every listing line={%
    \NotCopyable{\textcolor{red}{\lstshellstyle\$ }}%
  },
  after={\par\noindent},
}

\newcommand*{\shparambox}[1]{%
  \tcbox[on line, boxsep=0pt, left=1pt, right=1pt,
         top=0pt, bottom=0pt, toprule=0pt, bottomrule=0pt,
         arc=0pt, outer arc=0pt, colback=shback, colframe=shframe]{%
    \lstshellstyle\strut #1%
  }
}%

\makeatletter
\@ifclassloaded{beamer}{%
  \setenumerate[1]{
    label=\protect\usebeamerfont{enumerate item}%
          \protect\usebeamercolor[fg]{enumerate item}%
          \insertenumlabel.%
  }
  \setitemize{
    label=\protect\usebeamerfont*{itemize item}%
          \protect\usebeamercolor[fg]{itemize item}%
          \protect\usebeamertemplate{itemize item}%
  }
}{                              % not beamer:
  \newlist{shparamdescr}{description}{4}
  \setlist[shparamdescr]{
    format=\mdseries\shparambox,
    nosep,
    left=0pt,
    labelindent=0pt,
    labelwidth=*,
  }
}
\makeatother

\lstMakeShortInline[%
  language=sh,
  breaklines,                   %allow breaks in inline code.
]@%
\newcommand\inlinesh[1]{\lstinline[language=sh,breaklines]!#1!}%

\newminted{text}{%
  tabsize=2,
  breaklines,
  fontsize=\footnotesize\smaller,
}
\BeforeBeginEnvironment{textcode}{\begin{logbox}}%
\AfterEndEnvironment{textcode}{\end{logbox}\noindent}%

\usepackage{listings}

\excludecomment{leaveout}
\title{Tutorials on Testing Neural Networks}
\author{Nicolas Berthier, Youcheng Sun, Wei Huang, \\ Yanghao Zhang, Wenjie Ruan, Xiaowei Huang}
\date{March 31st, 2021}
\publishers{University of Liverpool}

\usetikzlibrary{external}
\AtBeginDocument{\tikzexternaldisable} %not by default

\tcbuselibrary{external}
\tcbset{external/prefix=tikz-cache/tcb-}
\tcbEXTERNALIZE

\begin{document}

\maketitle

\section*{Abstract}

Deep learning achieves remarkable performance on pattern recognition, but can be vulnerable to defects of some important properties such as robustness and security.
This tutorial is based on a stream of research conducted since the summer of 2018 at a few UK universities, including the University of Liverpool, University of Oxford, Queen's University Belfast, University of Lancaster, University of Loughborough, and University of Exeter.
This research is supported by the Defence Science and Technology Laboratory (Dstl), UK. Dr Xiaowei Huang is the PI of these projects. 

The research aims to adapt software engineering methods, in particular software testing methods, to work with machine learning models.
Software testing techniques have been successful in identifying software bugs, and helping software developers in validating the software they design and implement.
It is for this reason that a few software testing techniques -- such as the MC/DC coverage metric -- have been mandated in industrial standards for safety critical systems, including the ISO26262 for automotive systems and the RTCA DO-178B/C for avionics systems.
However, these techniques cannot be directly applied to machine learning models, because the latter are drastically different from traditional software, and their design follows a completely different development life-cycle.

As the outcome of this thread of research, the team has developed a series of methods that adapt the software testing techniques to work with a few classes of machine learning models.
The latter notably include convolutional neural networks, recurrent neural networks, and random forest.
The tools developed from this research are now collected, and publicly released, in a GitHub repository: \url{https://github.com/TrustAI/DeepConcolic}, with the BSD 3-Clause licence.

This tutorial is to go through the major functionalities of the tools with a few running examples, to exhibit how the developed techniques work, what the results are, and how to interpret them.

\subsection*{Contributors}

The following is a list of authors who have contributed to the software and the tutorial (their names are ordered alphabetically): Nicolas Berthier, Wei Huang, Xiaowei Huang, Wenjie Ruan, Youcheng Sun, Yanghao Zhang.

\setcounter{tocdepth}{1}
\tableofcontents

\chapter{Introduction}

\paragraph{Safety-critical Systems}
%\itodo{Basic introduction to Safety-critical systems}
are those systems whose failure could result in loss of life, significant property damage, or damage to the environment. Aircraft, cars, weapons systems, medical devices, and nuclear power plants are the traditional examples of safety-critical software systems.
The failures of safety-critical software systems have brought some companies and their software development practices to the attention of the public, \eg Boeing's two 737 Max groundings \cite{boeing737max} and the crash of Uber's self-driving car \cite{uber-car-crash}.

%\itodo{Random gibberish on system and specification: temporary.}
In such systems, one ideally seeks an absolute certainty that the system meets its specification.
In the case of software systems, this notably entails that the software does not contain any bug, or other internal sources of failure.
In the worst case, if such a failure is permitted to occur, one wants to ensure that it is properly detected, and that appropriate counter-measures are deployed. Please refer to \cite{HUANG2020100270,10.1145/3459086.3459636} for reviews on the recent progress on techniques for the safety and trustworthiness of neural networks. 

%\itodo{\emph{exhaustiveness}; and requirement that every behaviour of
%  the system meets a portion of the specification.}

\paragraph{Testing for Safety-critical Systems}

%\itodo{Specifications for safety-critical systems are required to meet
%  certain \emph{exhaustiveness} criteria to form a consistent basis
%  for safety assurance.}
Testing is still the primary approach in industry for increasing the confidence in software products and services.
Obviously, tests are non-exhaustive in every relevant practical cases (and almost by definition).
Yet, that does not prevent them from being pervasive even in safety-critical contexts where formal verification techniques are available to assess the correct behaviours of systems.
The main underlying reasons are twofold:
\begin{enumerate}[(i),nosep]
\item\label{enum:tests-assess-that-specification-is-realistic}%
  the successful design of test beds serves as an assessment that the specification is realistic (\ie it can be translated into a set of test cases);
\item\label{enum:tests-are-executable-specifications}%
  test beds are often the only ``executable specification'' that designers have access to in the implementation of the system itself.
  As such, they help
  % \item exercising the system in this way is a useful device to
  increase the level of confidence that the design meets its specification.
\end{enumerate}

We can draw some interesting observations from the above two points and the usual development process for ML systems.
Indeed, the core assumption of such systems, which states that expected system behaviours can be learned from data via training, exactly mirrors point~\ref{enum:tests-assess-that-specification-is-realistic}.
Furthermore, point~\ref{enum:tests-are-executable-specifications} refers to the iterative and empirical process for designing ML systems, that often starts from a bare neural network architecture, and evolves according to its results in meeting various requirements---\eg prediction accuracy, robustness to adversarial examples---via a series of architectural refinements.

Unfortunately, the parallel stops here: one step that is missing above for ensuring the safe operational use of a safety-critical system consists in obtaining a formal guarantee that it meets its specification, and this has no equivalent in the domain of machine learning systems.
One notable reason is that existing code coverage criteria for testing conventional software cannot realistically be applied to machine learning.

\paragraph{Robustness Concern}

One of the desirable properties that ML systems should satisfy is \emph{robustness}.
This concern was first revealed by~\citet{szegedy2013intriguing} when they showed that machine learning (ML) models, and in particular deep neural networks, may be vulnerable to \emph{adversarial attacks}, where small input perturbations cause the models to behave in an unintended way.
% Moreover, it is not possible to rely on a set of training data only
% to specify such systems for we cannot guarantee its completeness.
% We can, however, draw several interesting observations from these
% (dis)similarities.
% First,
%
% At last, an optimistic view might even extrapolate up to the
% far-fetched conclusion that an ML-enables system might be capable of
% providing an (intelligible) formal guarantee about its safe
% behaviours.\todo{Very very very far-fetched ;)}
%
% \begin{leaveout}
% \nb[inline]{Mention search for \emph{adversarial} examples to explore \emph{local robustness} property.}
% The works covered in this tutorial were partly fuelled by the discovery by~\citet{szegedy2013intriguing} that machine learning (ML) models, and in particular deep neural networks, may be vulnerable to \emph{adversarial attacks}, where small input perturbations cause the models to behave in an unintended way.
In the case of a model that achieves a \emph{classification task}, for instance, such an unintended behaviour is typically a misclassification of the perturbed input.
More generally, we can say that a \emph{classifier} is non-robust if it associates two very similar inputs (\eg two images that differ by very few pixels or brightness) with two distinct labels.

% \paragraph{On the Relevance of Coverage-guided Testing in Safety-critical Context}

% As is the case for traditional software, the relevance of a coverage criterion \C builds upon the \emph{assumption} that a neural network that correctly fulfils its function must pass a set of tests that meets \C.
% Furthermore, that a software ``passes a set of tests'' is also subject to varying definitions for the \emph{test oracle}.
% Likewise, this kind of assumption bears heavy implications on the potential and relevance for using coverage-guided testing for the validation of neural networks in the context of safety-critical systems.

% Mention search for \emph{adversarial} examples to explore \emph{local robustness} property.

% \paragraph{Testing for Robustness}
\paragraph{Coverage-guided Testing}

% All the tools covered in this tutorial focus on \emph{classification tasks}, although the implemented algorithms can be adapted to handle models dedicated to regression tasks.
% For instance, this can be done via the specification of interval bounds on the output error, to define a discrimination criterion so as to determine incorrect regression results.
%
% . classification obtained by the tested model when it is fed with the generated input, against the known classification of an input that is close to the generated one.
% This way, generated tests that do not pass this oracle are adversarial examples as they
Most of the tools we shall present address the validation of robustness via a series of \emph{coverage-guided testing} algorithms, where a predefined quantification of model behaviours is used to guide the generation of new tests.
To detect an adversarial example that demonstrates a violation of the aforementioned robustness property, the \emph{test oracle} typically compares the output of the model on a test \(x'\), against the output obtained on a test that is both ``close to'' \(x'\) and known to be correctly classified by the model.

From the above description, we can identify several ingredients that define how each tool operates:
\begin{enumerate}[(i),nosep]
\item the definition of coverage metrics and criteria (\eg structural, temporal, semantics), that guide the underlying test generation algorithm;
\item the approach for generating new tests in a way that increases coverage (\eg random mutations, symbolic analysis, gradient analysis);
\item the test oracle that detects violations of desired properties (\eg robustness).
\end{enumerate}
% In the list above, the principles underlying the oracle is common for most of the tools that we will present.
% Essentially,
In this tutorial, we will describe and illustrate the particular choices we have made for each one of these components in order to achieve coverage-guided testing for several families of ML models.
% We first review practical aspects .
We illustrate how to exercise each tool in practice, and rely on a common set of datasets to clearly distinguish the particularities of each approach.

\paragraph{Outline}

Practical aspects related to the installation and setup of the tools are given in Part~\ref{part:installation-n-setup}, and datasets are introduced in Part~\ref{part:datasets}.
We then consider each tool in turn, in their respective part:
\begin{description}[labelindent=1em,itemsep=1ex]
\item[\DeepConcolic] for testing Convolutional Neural Networks (CNNs) (Part~\ref{part:deepconcolic});
\item[\TestRNN] tests Recurrent Neural Networks (RNNs), and in particular Long-Short-Term Memory Models (LSTMs) (Part~\ref{part:testrnn});
\item[\EKiML] specifically addresses poisoning attacks and defence for Random Forests (Part~\ref{part:ekiml}); and
\item[\GUAP] seeks the discovery of spatial and/or additive universal adversarial attacks that fool an image classifier on a full set of images (Part~\ref{part:guap}).
\end{description}

\paragraph{What's not included?} 

In addition to the testing methods this tutorial (and the above-mentioned software packages) cover, there are verification methods such as \cite{HKWW2017,RHK2018,wicker2018feature,ruan2018global,LLYCH2018,wu2018game} 
that can provide the mathematical guarantees on whether a local robustness property is satisfied or not. Usually, the guarantees come with high computational complexity, and these methods do not scale to the neural networks that can be dealt with by testing methods. Moreover, this tutorial does not consider safety case construction, which as discussed in e.g., \cite{2020arXiv200305311Z,RAM2021} may utilise the results of verification and testing as evidence to build safety argument, and the verification and validation of systems where machine learning models are components such as \cite{9196932,9340720}.
%These include e.g., \cite{}. 

% \nb[inline]{As outline, review all tools, how they address the
%   aforementioned problems.}

\chapter{Installation \& Setup}
\label{part:installation-n-setup}

\section{Downloading the Software}

The software is publicly released on its GitHub page:

\begin{tcolorbox}[center, boxsep=0pt, left=1ex, right=1ex, width = 9cm, before upper=\strut, tcbox width = minimum center, box align=center, halign=center, colback=white, colframe=blue!50!black]
  \mbox{\url{https://github.com/TrustAI/DeepConcolic}}
\end{tcolorbox}

As the first of step of installation, the following commands can be executed to download the software:
%
% !setup download git
\begin{cmds}
git clone https://github.com/TrustAI/DeepConcolic.git -b tutorial
cd DeepConcolic
\end{cmds}
Throughout this document, we will give details on how to actually use the tools included in this directory (\ie \DeepConcolic, \EKiML, \testrnn, and \GUAP), % to achieve efficient testing
and explore their respective results.
We will assume that commands are executed \emph{from within the} @DeepConcolic@ \emph{directory}.

\begin{warn}
  Observe that the @git@ command above retrieves the \shparambox{tutorial} branch of the source code.
  This branch includes versions of the tools that perform basic checks to ensure the above assumption, \ie that executions take place from within the @DeepConcolic@ directory.
  % and that the @conda@ environment is properly setup.
  % The tools included in the @tutorial@ retrieved above perform some basic check to ensure that this is the case.
\end{warn}

\section{Setting up a Software Environment and Installing Dependencies}

Once the software is downloaded, we need to install an Anaconda through the link \url{https://docs.anaconda.com/anaconda/install/}.
We set up an Anaconda environment called @deepconcolic@ with the following commands:
% !setup conda create conda-activate
\begin{cmds}
conda create --name deepconcolic python==3.7
conda activate deepconcolic
\end{cmds}
We can now install the additional software dependencies within this environment by using the following commands:
% install -c conda-forge rdkit
% !setup conda pip conda-install pip-install
\begin{cmds}
conda install opencv nltk matplotlib
conda install -c pytorch torchvision
pip3 install numpy==1.19.5 scipy==1.4.1 tensorflow\>=2.4 pomegranate==0.14 scikit-learn scikit-image pulp keract np_utils adversarial-robustness-toolbox parse tabulate pysmt saxpy keras menpo patool z3-solver
\end{cmds}
% Note `pytorch' and `torchvision' are for GUAP.

\begin{warn}
  Similarly to the checks about the current directory, the tools provided in the \shparambox{tutorial} branch also ensure that the @conda@ environment is properly setup, \ie the \shparambox{conda activate deepconcolic} command has been executed.
\end{warn}

\section{Checking Installed Solvers}
\label{sec:check-inst-solv}

Many of the tools that we are covering in this tutorial internally rely on dedicated problem solvers that may need to be installed separately.
In particular, \EKiML and \DeepConcolic use SMT and LP solvers, respectively.

\paragraph*{SMT Solver for \EKiML: Z3}

\EKiML makes use of Z3 for solving \emph{Satisfiability Modulo Theories} (SMT) problems (this choice ensures appropriate performances of the tool, and is at the moment not configurable).
To ensure that this solver is available, we can run the following command and check that the printed list of solvers includes @z3@:
% !setup EKiML check z3
\begin{cmds}
python3 -m pysmt install --check
\end{cmds}
\begin{textcode}
…
Solvers: z3
…
\end{textcode}
% If required, @z3@ may be installed via the following command:
% % !setup EKiML install z3 silent
% \begin{cmds}
% python3 -m pysmt install --z3
% \end{cmds}

\paragraph*{LP Solver for \DeepConcolic}

Several concolic algorithms implemented in \DeepConcolic make use of a \emph{Linear Programming} (LP) problem solver to generate new test cases.
We % check what solvers are available with:
% % !setup check pulp solvers
% \begin{cmds}
% pulptest
% \end{cmds}
% and
check what solver is selected by default by \DeepConcolic with:
% !setup DeepConcolic check pulp solvers
\begin{cmds}
python3 -c 'from deepconcolic import lp; lp.pulp_check ()'
\end{cmds}
\begin{textcode}
Using TensorFlow version: 2.4.0
PuLP: Version 2.4.
PuLP: Available solvers: GLPK_CMD, PULP_CBC_CMD, COIN_CMD, PULP_CHOCO_CMD.
PuLP: COIN_CMD solver selected (with 10.0 minutes time limit).
\end{textcode}
Note the solvers listed above all work by writing LP problems into files on disk and launching a separate process (this is indicated by the ``@_CMD@'' suffixes).
Significant performance gains can be obtained for the concolic algorithms of \DeepConcolic by installing a solver that provides direct python bindings such as CPLEX\footnote{Please see \url{https://coin-or.github.io/pulp/guides/how_to_configure_solvers.html} for detailed instructions for installing and setting up additional solvers.}.

\section{Optional Sand\-boxing}
\label{sec:sandboxing}

The original purpose of most of the tools that we are covering in this tutorial is to support academic experiments with the proposed algorithms.
It is therefore legitimate to assume that unanticipated uses of these tools may inadvertently have unintended effects on your systems (such as, in the worst case, delete full directory hierarchies).
To prevent this from happening, we provide a means to protect your file-systems against these potentially harmful operations.
On Debian-based GNU/Linux systems (\eg Ubuntu), we suggest using \textsf{bubblewrap}\footnote{\url{https://github.com/containers/bubblewrap}, usually available as a distribution package named ``\texttt{bubblewrap}''.}
via a helper script that we provide to safely experiment with the tools.
The latter can be used by running the following command from within the main \DeepConcolic directory.
% !silent wrap bubblewrap
\begin{cmds}
HISTFILE=.histfile tuto-scripts/bwrap-shell.sh /bin/bash
\end{cmds}
The above command starts a @bash@ shell process (and any subsequently forked sub-process, launched commands) that essentially has no write access to any directory other than the current directory and @/tmp@.
It additionally sets the environment variable @HISTFILE@ in such a way that the history of typed commands is still made persistent after the wrapper terminates (since typically the root of the user home directory cannot be written by the newly launched @bash@ interpreter).
Note that the wrapper script we provide may fail if executed under a network-mounted directory hierarchy (\eg via NFS)% , like under some home
% directories
.

\section{Testing the Installation}

We can test if the software is correctly installed and ready to go by running a few commands such as
% !setup deepconcolic testrnn EKiML GUAP check
\begin{cmds}
python3 -m deepconcolic.main -h
python3 -m testRNN.main -h
python3 -m EKiML.main -h
python3 -m GUAP.run_guap -h
\end{cmds}
which should print usage messages for \DeepConcolic, \testrnn, \EKiML, and \GUAP.
% python3 -m EKiML.main --Dataset iris --Mode embedding --Embedding_Method black-box --Model forest

\chapter{Example Datasets% and Machine Learning Models
}
\label{part:datasets}
\label{cha:example-datas-mach}

We consider two datasets as running examples of this tutorial: Fashion-MNIST and UCI-HAR.

\section{Fashion-MNIST}
\label{sec:fashion-mnist}

Fashion-MNIST~\cite{xiao2017/online} is a dataset of Zalando's article images, consisting of a training set of \num{60 000} examples and a test set of \num{10 000} examples.
Each Fashion-MNIST example is a 28x28 grayscale image, associated with a label from 10 classes.
Although the images are in the same format as in the widespread handwritten digit recognition MNIST dataset, the classification task for Fashion-MNIST is deemed more complicated.

\vfill
\centerline{\includegraphics[width=.86\textwidth]{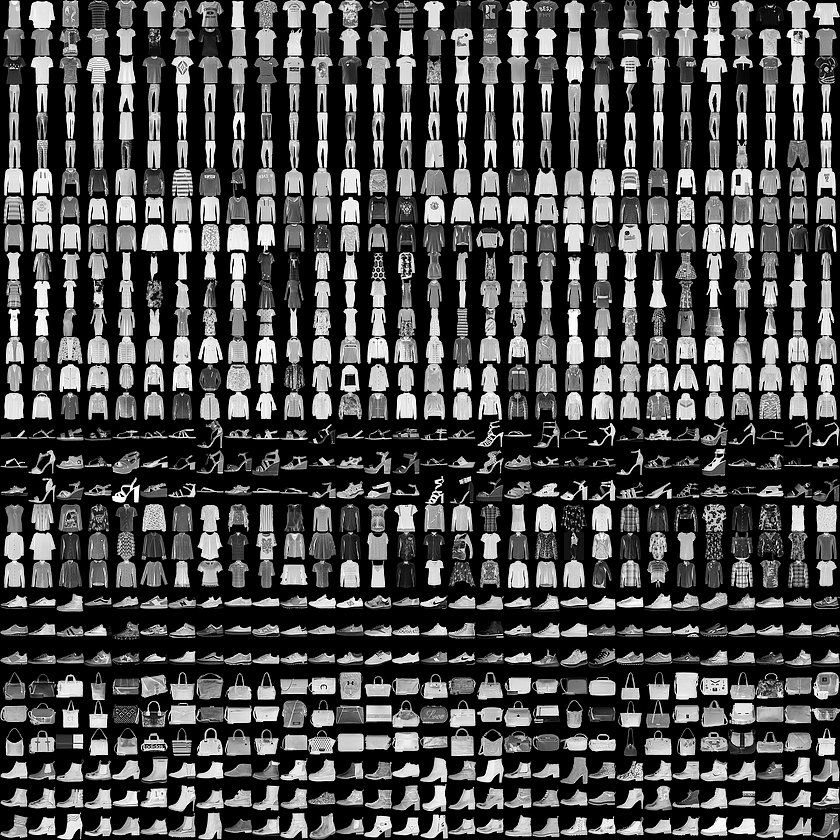}}

\section{UCI-HAR: Human Activity Recognition (Preprocessed)}
\label{sec:uci-har}

UCI-HAR database (further abbreviated HAR in the following) was built by \citet{anguita2013public-HAR} from the recordings of 30 subjects performing activities of daily living %(ADL)
while carrying a waist-mounted smartphone with embedded inertial sensors.
%This dataset version contains all the training and testing examples provided in the original data repository.
%
Each record in the dataset consists in a 561-feature vector with time and frequency domain variables, along with an associated activity label.
For each entry, the recorded variables include:
% For each record in the dataset, it is provided:
\begin{itemize}
\item triaxial acceleration from the accelerometer (total acceleration) and the estimated body acceleration;
\item triaxial angular velocity from the gyroscope.
\end{itemize}
The classification task at hand is to determine whether each sample corresponds to sensor readings obtained while the subject performed one of six activities: walking, walking upstairs, walking downstairs, sitting, standing, or lying down.

Unlike the other tools, \EKiML does not automatically download the data.
We therefore store the dataset on our server, so that it can be downloaded by running the following command:
% !download har dataset
\begin{cmds}
wget -P datasets https://cgi.csc.liv.ac.uk/~acps/datasets/UCIHARDataset.csv
\end{cmds}

% \section{Other Datasets}
% \label{sec:other-datasets}

% Some of the tools' functionalities are better illustrated using other datasets.
% For instance, in Part~\ref{part:guap} about \GUAP, we also consider CIFAR-10 and ImageNet as they provide more illustrative visual examples.
% These datasets do not need to be downloaded beforehand as they are retrieved automatically by the tool.

\chapter{Testing Convolutional Neural Networks}
\label{part:deepconcolic}
%(written by Nicolas and Youcheng)}

In this Section, we will explore how concolic executions and other test case generation algorithms implemented in \DeepConcolic can be used to support the design of ML-enabled safety-critical systems.
We will in particular focus on % deep and convolutional neural networks,
% and illustrate
the specific abilities of \DeepConcolic to achieve various test coverage criteria on \emph{feed-forward}, dense and/or convolutional neural networks.
We first give a succinct overview of these capabilities and explain how to exercise them in practice.
We then conduct experiments using our two running example datasets to both illustrate how \DeepConcolic can be used for testing, and the effects of various selected coverage criteria. Details about the DeepConcolic are reported in \cite{sun2018concolic,sun2018testing,SHKSHA2019}.

\section{Overview of \DeepConcolic}
\label{sec:overv-deepc}

Let us first briefly present the general working principles and purposes of \DeepConcolic.

\begin{figure}
  \centering
  % -*- coding: utf-8 -*-
%\tikzexternalenable%
\begin{tikzpicture}[>={Stealth[scale=.8]}, node distance = 1cm and 1cm]
  \tikzset{every node/.style={%
      rectangle,
      align = center,
      inner sep = 1ex,
    }};
  \tikzstyle{engine-component} = [draw, double];
  \node (raw) {raw\\ data};
  \node [right = of raw] (suite) {test\\ suite};
  \node [above = of raw] (dnn) {DNN};

  \node [right = of suite, engine-component] (crit) {\strut Test Criterion};
  \node [above = of crit, engine-component] (analyzer) {\strut Search Engine};
  \node [above = of analyzer, draw, circle, inner sep = 2pt] (oracle) {\strut Oracle};
  \node [right = of crit, xshift = 1ex] (report) {coverage\\ report};
  \node at (oracle -| report) (adv) {adversarial\\ examples};
  % \node [above = 3mm of oracle] {\DeepConcolic Loop};

  \draw [->] (raw) to (suite);
  \draw [->] (suite) to (crit);
  \draw [->] (crit) to node [right] (t) {\smaller\mdseries test target} (analyzer);
  \draw [->] (crit) to (report);
  \draw [->] (dnn) to (analyzer);
  \draw [->] ([xshift=-4mm]dnn -| analyzer.west) |- ([yshift=2mm]crit.west);
  \draw [->] (analyzer) to node [right] (x') {\smaller\mdseries new test} (oracle);
  \draw [->] (oracle) to (adv);
  \draw [white,line width=1.6mm,shorten <= 2mm] (oracle) -| (suite);
  \draw [->] (oracle) -| (suite);

  \coordinate (xx) at ([xshift=-4mm]dnn -| analyzer.west);
  \node [draw, dashed, rectangle, fit={(crit)(analyzer)(xx)(x')(oracle)}, inner sep = 3mm] {};
  \node [above left = 2mm and 5mm of oracle, anchor = east, fill = white, fill opacity = .9, text opacity = 1] {\DeepConcolic Loop};
\end{tikzpicture}
\hrule width 0pt              %dummy spacer to end vbox earlier
 \par
  \begin{adjustwidth}{1.5cm}{1.5cm}
    \medskip
    \centerline{\hbox to .82\hsize{\hrulefill}}
    \medskip
    \itshape
    \begin{description}[format=\upshape\mdseries]
    \item[Test Criteria:]
      \begin{itemize*}
      \item Structural (neuron-based)
      \item High-level (\LF-based)
      \end{itemize*}
    \item[Search Engines (Concolic):]
      \begin{itemize*}
      \item Linear programming
      \item Pixel\-wise optimisation
      \end{itemize*}
    \item[Search Engines (Other):]
      \begin{itemize*}
      \item Gradient analysis
      \item Fuzzing
      \end{itemize*}
    \end{description}
  \end{adjustwidth}
  \caption{Overview of \DeepConcolic's Architecture}
  \label{fig:deepconcolic-overview}
\end{figure}
We give an overview of \DeepConcolic in \figurename~\ref{fig:deepconcolic-overview}.
\DeepConcolic takes as input a \emph{feed-forward neural network} \N and some raw data.
The latter includes both a \emph{test dataset} \Xtest associated with classification labels \Ytest, along with a \emph{training dataset} \Xtrain (with associated labels \Ytrain) that has been used to train \N (\ie compute its parameters).

\subsection{Coverage-guided Testing}
\DeepConcolic then attempts to generate \emph{new test inputs} according to a \emph{concolic testing} algorithm, that it inserts into \Xtest so that the full test dataset fulfils a \emph{predefined coverage criterion}.
This process achieves coverage more efficiently than a randomised generation of new inputs by involving a \emph{search engine} that is dedicated to a generate new inputs% that fulfil the sought after coverage
% criterion
: the search engine takes a given \emph{test target} \(t\), which is selected in such a way that any new test that meets (\ie satisfies) \(t\) improves the coverage criterion over \Xtest; \ie \(t\) is not met by \Xtest.
% \(t\) is typically expressed in terms of the network's structure (\eg
% layers, neurons) and
The role of the search engine is to generate a \emph{new test input} \(x'\) that achieves \(t\).
\(x'\) is then checked against an \emph{oracle}, that determines whether it consists in a sufficiently realistic new input, and then if it is correctly classified by \N: \(x'\) is inserted into \Xtest if it passes the oracle, or rejected otherwise.
Further, any test inserted in \Xtest that is not classified correctly by \N is an \emph{adversarial example}.
The algorithm progresses towards achieving the coverage criterion as long as the search engine is guaranteed to produce inputs that meet the test targets.

\subsection{Main Components of \DeepConcolic}
From the above description, one can identify several main components that drive \DeepConcolic's operations:

\paragraph{Test criterion}%
The test criterion embodies the objective of the testing algorithm, via the definition of the coverage criterion.
Just like coverage is often defined in terms of statements or lines of code in the case of traditional software, for neural networks we consider criteria that are typically expressed in terms of the network's structure (\eg neurons, layers).
Coverage criteria in \DeepConcolic can be divided into the two following categories:
\begin{subdesc}
\item[Structural] criteria typically refer to values pertaining to individual \emph{neurons};
\item[High-level] criteria take a step towards covering the \emph{semantic features} that have been learned by \N% with the help
  % of a linear dimensionality reduction of neuron outputs induced by
  % training data.
  ;
\end{subdesc}
We will give more details about each criterion below.
\paragraph{Oracle}%
To vet a new input as legitimate (\ie sufficiently close to a reference input), and check whether it is correctly classified by \N, the oracle relies on a given \emph{norm} and, optionally, on a (series of) \emph{post-filters}:
\begin{subdesc}
\item[Norms]%
  are distance metrics that can be used to assess the similarity of two given inputs.
  In \DeepConcolic, the norm may be based on the Chebyshev distance \(L_∞\), that uses the maximum absolute difference between any input feature (\ie colour of pixel, input scalar) as a measure of distance, or the \(L_0\) ``norm'', that simply counts the number of input features that differ.
  We denote with \LInf x y the distance between two inputs \(x\) and \(y\) \wrt the \(L_∞\) norm (and similarity for \(L_0\)).

  The oracle rejects any input \(x'\) \st \(\LInf x {x'} > \dthr\), for some given threshold \dthr.
\item[Post-filters]%
  take a generated input and decide whether it satisfies some requirements that cannot be expressed by specifying a norm.
  Such filters can for instance be useful to deal with cases where some input features have particular semantics, as opposed to pixels of images that may just take their values in a fixed range and do not need to obey global well-formedness constraints.
  We elaborate more on a typical use for post-filters in Section~\ref{sec:oracle-augm-postfilters} below, and exemplify its use in Section~\ref{sec:human-activ-recogn};
\end{subdesc}
If \(x'\) passes all the above filters, the oracle compares the classification labels assigned by \N to \(x\) and \(x'\): if the labels differ, then \(x'\) is an adversarial example.

\paragraph{Search engines}%
The search engine goes hand in hand with the selected criterion, although several engines may be available for each given criterion, depending on the underlying analysis technique, as well as the norm selected.
\DeepConcolic includes three distinct families of search engines:
\begin{subdesc}
\item[Concolic engines] make use of symbolic encoding of (a subset of) the neural network.
  Such an engine takes both the test target \(t\) and a \emph{candidate input} \(x\) to construct the encoding in such a way that its solution forms a new input \(x'\) that is close to \(x\) \wrt the considered norm.
  Most concolic engines construct their symbolic encoding as a \emph{Linear Programme} (LP), and are therefore restricted to using a norm that is also a \emph{linear metric} for this encoding to be possible; in \DeepConcolic, only the \(L_∞\) norm satisfies this property.
  Other concolic engines use ad-hoc global optimisation approaches that can be used to support non-linear norms like \(L_0\).
\item[GA-based search engines] operate differently from their concolic counterparts, as they approach the problems via \emph{Gradient Analysis} (GA).
  In \DeepConcolic, the latter is delegated to an attacker (\ie a generative adversarial network) based on the network \N under consideration, that is then used to efficiently generate new inputs via gradient analysis %descent% from
  % candidate inputs% in such a back-propagating the targeted
  % changes% , and
  % % select those for which on neuron outputs to construct new inputs
  .
  By virtue of the latter, these engines are more likely to generate new examples that are actual adversarial examples. %\nb{Is that true?}
\item[Naive engines] do not target specific coverage criteria, and naively generate many new inputs by randomly mutating legitimate inputs%
  % by involving specially crafted search
  % algorithms% (for L0 norm \& neuron coverage)
  .
\end{subdesc}

% \paragraph*{Global \vs Rooted Search}

Some search engines only operate based on a given candidate input \(x\) in addition to the test target \(t\): we say that these engines are \emph{rooted} as they only \emph{derive} new inputs---thereby constructing a hierarchy of inputs---, whereas engines that operate by scanning a whole set of reference inputs are said to perform a \emph{global search}.
All concolic engines are rooted, whereas GA-based ones are global.
This notably means that, in the case of a rooted search, some \emph{heuristics} need to be employed in the search of a suitable pair \((t, x)\); this is done by the test criterion component.
Furthermore, the size of the initial test dataset, that we shall denote \(\Xtest_0\), also has implications on the ability of said heuristics to find suitable candidate inputs.

% \subsubsection{Norm Parameters}
% \label{sec:norm-parameters}

\subsection{Norm Parameters for Concolic Search Engines}
\label{sec:norm-param-conc}

In addition to the threshold factor \dthr used by the oracle to assess the plausibility of generated inputs, a pair of parameters pertaining to the norm are available to tune the exploration of the norm ball around each candidate input \(x\) when selecting a concolic engine:
\begin{mathpardesc}[nosep]
\item[\dminHard] gives a lower bound for \LInf x {x'};
\item[\dminNoise] introduces some uniform random noise on the lower-bound to help further explore the input space.
\end{mathpardesc}
Then, each time an LP problem is constructed, a lower bound for \LInf x {x'} is drawn uniformly in the range \([\dminHard, \dminHard + \dminNoise]\): this mechanism allows users to control the minimum changes to candidate input features that can be performed at each step of the test case generation algorithm.

\subsection{Oracle Augmentation with Post-filters}
\label{sec:oracle-augm-postfilters}

\DeepConcolic features a means to augment the capabilities of oracles by using a post-filter that relies on an estimation of \emph{Local Outlier Factors}~\citep{Breunig2000LOF}.
% As of now, the available filters rely on an % unsupervised
% estimator dedicated to
% % First, the @--filters LOF@ arguments triggers the initialisation of an
% % estimator that
% rely on the computation of \emph{Local Outlier
%   Factors}~\citep{Breunig2000LOF}.
When enabled, this device is used in addition to the oracle mechanism described previously, to detect and filter out generated inputs that can be considered outliers \wrt the raw training data.
The underlying estimator is initialised with a set \(X_{\mathtt{LOF}}\) of inputs using a specialised search tree that can be used to identify the \(k\)-nearest neighbours in \(X_{\mathtt{LOF}}\) that are the closest to any new input \(x\) \wrt some configurable distance metric.
The local outlier factor of \(x\) is then computed based on the density of its \(k\)-nearest neighbours, and this factor is used to determine the plausibility of \(x\) using a predefined threshold.
% that lies above some baseline threshold value used by the
% oracle to filter out implausible inputs.

Of course, the choice of the distance metric is of paramount importance for this mechanism to be relevant and act as a useful oracle.
By default the LOF-based filter uses the \emph{cosine distance} as a measure of distance between two inputs.
This distance typically gives an appropriate indication of the relative similarity between two 1-dimensional data vectors, as it only considers the angle between the two normalised vectors and disregards the magnitude of each one of their respective components.
% the respective features for 1-dimensional data vectors.

\subsection{Additional Assumptions \& Requirements}
\label{sec:dnn-arch-requ}

As of now, \DeepConcolic can only be used for testing feed-forward classifiers; \ie neural networks that achieve regression functions are not supported.
Note, however, that the overall approach adopted by \DeepConcolic can often be extended to such estimators.
For instance, this can be done via the specification of interval bounds on the output error, to define a discrimination criterion so as to determine incorrect regression results.

\section{Coverage Criteria for Convolutional Neural Networks}
\label{sec:coverage-criteria}

We can now review the various coverage criteria available in \DeepConcolic.
Depending on the semantics of the underlying coverage metrics, they can currently be partitioned into \emph{structural criteria} on one side, and \emph{high-level criteria} on the other side.

\subsection{Structural Criteria}

The structural coverage criteria, and associated search engines implemented in \DeepConcolic were originally designed and presented by~\citet{SHKSHA2019}.
These notably include \emph{neuron coverage} (NC), which basically counts the amount of neurons activated by the test dataset---a neuron is said \emph{activated} if the value output by its (ReLU) activation function is positive.
Other structural variants that are available are inspired by the \emph{modified condition/decision coverage} (MC/DC) used in software testing: in \DeepConcolic, this category of \emph{combinatorial coverages} includes sign-sign coverage (SSC) ~\citep{sun2018concolic,sun2018concolicb}, that seeks to witness every combination of neuron activations in each pair of successive layers of the network.

\subsection{High-level Criteria}
\label{sec:high-level-criteria}

The family of high-level criteria available in \DeepConcolic have been investigated by~\citet{berthier2021abstraction}.
They are defined based on a \emph{Bayesian Network abstraction}, which constitutes an abstraction of all behaviours of the neural network subject to a given dataset.
Every node in the BN is attached to a particular layer, and represents a component from a low-dimensional space that efficiently explains all combinations of neuron outputs for the layer when the neural network is subject to the training data \Xtrain.
This is achieved by means of a \emph{linear dimensionality reduction technique} (such as Principal Component Analysis---PCA--- or Independent Component Analysis---ICA), and the underlying assumption is that each extracted dimension for a hidden layer captures a high-level ``\emph{\LF}'' that has been learned by the neural network under consideration.
Each node of this BN represents a random variable that ranges over a finite set of \emph{intervals} that partitions the \LFs, and the BN represents a joint probability distribution for all the individual combinations of intervals for the \LFs.

Bayesian inference is used to compute the probabilities in the BN based on the neuron outputs induced by the test dataset \Xtest.
As a result, % combinations of
node values that appear with a high probability according to the BN represent behaviours that are well tested by \Xtest; conversely,
% combinations
node values that are rare \wrt the BN are not tested (enough) by \Xtest.
Then, a test dataset satisfies a high-level BN-based criterion if no
% combination of
\LFI is rare \wrt the BN: this defines the criterion that we call the \emph{Bayesian-network Feature Coverage} (BFC).

The BN can alternatively be seen as a set of conditional probability tables, that we can use to define an additional criterion that accounts for the \emph{causality assumption} underling the layered nature of the neural network, which states that the set of outputs at a particular layer depends on the set of outputs at a preceding layer.
Indeed, the conditional probability tables give, for each \LFI pertaining to hidden or output layers, whether some test in \Xtest simultaneously exhibits each combination of intervals at the preceding layer.
% expressed how an interval pertaining to a \LF extracted for a (hidden
% or output) layer is associated with its probability conditioned to a
% combination of all \LFIs at the preceding layer.
% These tables can therefore be used to further leverage a
% \emph{causality assumption} which states that the set of outputs at a
% particular layer depends on the set of outputs at a preceding layer.
This defines a criterion that we call the \emph{Bayesian-network Feature-dependence Coverage} (BFdC).

\subsubsection{Parameters Underlying the BN Abstraction}

The main parameters that drive the construction of the BN abstraction defined above relate to the choice of a linear feature extraction technique, along with the creation of \LFIs.

Regarding linear feature extraction, \DeepConcolic supports both PCA and ICA.
Both techniques take an Integer parameter \(|Λ_i|\) that describes, for each layer \layer i of the network, the number of components to extract from the set of neuron activations at the layer; \ie \(|Λ_i|\) corresponds to the number of \LFs (and thus BN nodes) pertaining to \layer i.

Regarding the discretization of each \LF, several strategies are available as well.
Basic strategies include ``uniform'' and ``quantile'', that each take the desired amount of intervals for each \LF as a parameter.
The difference between the two lies in that the former uniformly partitions the \LF segment that contains all points of the (projected) training dataset, whereas the latter computes interval boundaries so as to evenly spread the dataset in the resulting intervals.

% \begin{redenv}
%   Note that, contrary to
% For technical reasons, the sequential DNNs for which the BN-based
% approach that is implemented in \DeepConcolic is able to generate new
% test cases must satisfy some basic structural requirements.
% First, except for the output layer, every activation function in the
% network must belong to a layer on its own.
% The main reason behind this limitation lies in that the components of
% the BN representation is computed based on the values output by
% neurons before they are fed into activation functions.
% \end{redenv}

\section{Practical Aspects in using \DeepConcolic}
\label{sec:pract-aspects-using}

The following command prints a help message for \DeepConcolic:
% !deepconcolic usage
\begin{cmds}
python3 -m deepconcolic.main -h
\end{cmds}
\DeepConcolic always requires the specification of a dataset (given with \shparambox{--dataset}), a trained neural network to test (given with \shparambox{--model}, that accepts trained networks in Keras H5 format\footnote{See \url{https://www.tensorflow.org/guide/keras/save_and_serialize\#keras_h5_format}}), and a directory where it will place all the output files it generates (with \shparambox{--outputs}).
To perform testing, \DeepConcolic also requires the specification of a criterion and a norm (with arguments \shparambox{--criterion}, \shparambox{--norm}); the search engine is then selected automatically from the two latter.
The following command-line arguments may additionally be given to specify all further parameters mentioned above.
\begin{shparamdescr}
\item[--init $|\Xtest_0|$] specifies the size of \(\Xtest_0\) (1 by default);
\item[--max-iterations $N$] gives the maximum number of iterations of the algorithm;
\item[--norm-factor \dthr] specifies the norm distance above which generated inputs are rejected as non-plausible by the oracle (its default value is \(\dthr = \frac 1 4\));
\item[--lb-hard \dminHard] and \shparambox{--lb-noise \dminNoise} respectively give parameters for tuning the exploration of the norm ball by LP-based concolic search engines (see Section~\ref{sec:norm-param-conc}).
  Default values are \(\dminHard = \frac 1 {255}\) for image datasets (\(\frac 1 {100}\) otherwise) and \(\dminNoise = \frac 1 {10}\);
\item[--filters LOF] enables oracle augmentation using the LOF-based post-filter with cosine distance (see Section~\ref{sec:oracle-augm-postfilters}).
\end{shparamdescr}
% \begin{description}[nosep]
% \item[\(|\Xtest_0|\)] --- %
%   The desired size of \(\Xtest_0\) can be given via the command-line
%   argument \shparambox{--init} (it is 1 by default);
% \item[\dthr] --- %
%   The norm distance above which generated inputs are rejected as
%   non-plausible by the oracle can be specified using the
%   command-line argument \shparambox{--norm-factor} (its default
%   value is \(\dthr = \frac 1 4\));
% \item[\dminHard and \dminNoise] (LP-based concolic search engines
%   only) --- %
%   These parameters can be specified with command-line arguments
%   \shparambox{--lb-hard} and \shparambox{--lb-noise}, respectively.
%   Default values are \(\dminHard = \frac 1 {255}\) (for image
%   datasets, \(\frac 1 {100}\) otherwise) and \(\dminNoise = \frac 1
%   {10}\);
% \item[\(N\)] --- %
%   Optionally, a maximum number of iterations of the algorithm, which
%   also gives an upper-bound to the amount of generated tests, can be
%   specified via \shparambox{--max-iterations}.
% \end{description}

\paragraph{Selecting Covered Layers}

Coverage criteria in \DeepConcolic are measured in terms of the values output by neurons of \emph{covered layers} (\ie the activations of these neurons).
Although the tool automatically discovers relevant layers based on the selected criterion, it is advisable to manually specify a list of covered layers to avoid ambiguities.
This can be achieved by giving a list of layer names as argument to \shparambox{--layers}, \eg %
@--layers activation_1 activation_2@.
Any supported layer can be specified when targeting a high-level coverage criterion: \ie dense, convolutional, max-pooling, ReLU activation---or even dropout, flatten or reshape layers.
For technical reasons, however, only layers that output values via a ReLU activation function may be considered by structural criteria.

\paragraph{Miscellaneous}
\DeepConcolic accepts several additional useful flags:
\begin{shparamdescr}
\item[--save-all-tests] Dump every generated test (images, etc); without this flag the tool only saves adversarial examples;
\item[--setup-only] Instruct the tool to terminate right before starting the main \DeepConcolic loop: use this to check whether command-line arguments are appropriate to the considered models;
\item[--rng-seed] Specifies a seed for the internal random number generator: this can be used to obtain reproducible executions.
  Note, though, that this holds up to the non-determinism induced by underlying tools used by search engines, such as LP solvers.
\end{shparamdescr}

\subsubsection{Parameters \& Companion Tool for BN-based Criteria}

The specification of parameters for constructing the BN abstraction for high-level coverage described in Section~\ref{sec:high-level-criteria}, can be done via a YAML file given as argument to \shparambox{--dbnc-spec}.
We refer to the file @dbnc/example.yaml@ in the source code repository for a detailed and illustrated description of each parameter.
For the purpose of simplifying this tutorial, however, we will make use of a companion tool called \dbnabstr provided as part of \DeepConcolic.
This tool can notably be used to pre-compute BN abstractions and dump them into a file.
The latter can directly be given to \DeepConcolic for defining BN-based criteria using the command-line option \shparambox{--bn-abstr}.

The creation of an abstraction is done via
% !silent dbnabstr
\begin{cmds}
python3 -m deepconcolic.dbnabstr --dataset <Dataset> --model <Model> create <abstraction.pkl> ...
\end{cmds}
where \shparambox{--dataset} and \shparambox{--model} give the dataset and model in a similar way as for \DeepConcolic, and @<abstraction.pkl>@ is a @.pkl@ file where the abstraction is to be saved.
The main command-line parameters that drive the abstraction are:
\begin{shparamdescr}
\item[--layers] As for the main \DeepConcolic tool, this gives the DNN layers to consider for measuring coverage;
\item[--feature-extraction] In \{@pca@, @ipca@, @ica@\} (the default being @pca@), this parameter specifies the dimensionality reduction technique to use for linear feature extraction (@ipca@ is a memory-efficient, incremental version of @pca@);
\item[--train-size] How many samples of the training dataset to use for feature extraction (the default is all;
\item[--num-features] Number of extracted features for each layer considered (default is 2);
\item[--num-intervals] Number of intervals for partitioning for each extracted feature (default is 2);
\item[--extended-discr] Whether to compute intervals that partition the full domain of each \HF (\ie \(\mathbb{R}\)) instead of their respective range of values observed during training.
\end{shparamdescr}

% \section{Case Studies}
% \label{sec:pres-case-stud}
\medskip

We will now illustrate how to use \DeepConcolic to support the validation of our neural networks.
We will particularly focus on the intents and effects of the testing algorithms when applied using various norms and criteria.

\section{Example: Fashion-MNIST}
\label{sec:deepconcolic-fashion-mnist}

We start with the Fashion-MNIST image classification task for it provides results that can easily be visually inspected.
They show the specificity of coverage criteria, and give hints about the behaviours of \DeepConcolic's coverage-guided testing algorithms.

\subsection{Training or Downloading Models}

For ease of use and experimentation, \DeepConcolic's source code includes a script that enables one to construct and train two CNNs for this dataset.
It saves models under @/tmp@, and can be executed from within the repository's root directory via
% For instance, we provide a script at train a DNN model for this
% dataset with
the following command (note this script automatically downloads the dataset from the Internet):
% !train deepconcolic fashion-mnist dense medium large
\begin{cmds}
python3 -m deepconcolic.gen_fashion_mnist
\end{cmds}
Generated models can then be copied from @/tmp@ into the @saved_models@ directory (to be created, if necessary).
Alternatively, one can download pre-trained models from our server:
% !silent download deepconcolic fashion-mnist dense medium large
\begin{cmds}
wget -P saved_models https://cgi.csc.liv.ac.uk/~acps/models/fashion_mnist_medium.h5
wget -P saved_models https://cgi.csc.liv.ac.uk/~acps/models/fashion_mnist_large.h5
\end{cmds}
The commands above download two pre-trained models into the directory ``@saved_models/@'', where the model ``@fashion_mnist_medium.h5@'' is a medium-sized model whose layers are listed in \tablename~\ref{tab:cnn-fashion-mnist-medium}, and  ``@fashion_mnist_large.h5@'' provides a slightly larger model with more pairs of convolutional/max-pooling and a dropout layer for further experimental investigations.
\begin{table}
  \centering\smaller
  \begin{tabular}{clrr} \hline
    Layer      & Name \& Function specification           & Output shape & \#parameters \\ \hline
    \layer 0   & \texttt{conv2d} (convolutional)          & 26 × 26 × 32 & 320                           \\
    \layer 1   & \texttt{activation} (ReLU)               & 26 × 26 × 32 & 0                             \\
    \layer 2   & \texttt{max\_pooling2d} (max-pooling)    & 13 × 13 × 32 & 0                             \\
    \layer 3   & \texttt{conv2d\_1} (convolutional)       & 9 × 9 × 64   & \num{51 264}                  \\
    \layer 4   & \texttt{activation\_1} (ReLU)            & 9 × 9 × 64   & 0                             \\
    \layer 5   & \texttt{max\_pooling2d\_1} (max-pooling) & 4 × 4 × 64   & 0                             \\
    \layer 6   & \texttt{flatten} (flat)                  & 1024         & 0                             \\
    \layer 7   & \texttt{dense} (dense)                   & 100          & \num{102 500}                 \\
    \layer 8   & \texttt{activation\_2} (ReLU)            & 100          & 0                             \\
    \layer 9   & \texttt{dense\_1} (dense)                & 10           & \num{1 010}                   \\
    \layer{10} & \texttt{activation\_3} (softmax)         & 10           & 0                             \\
    \hline
  \end{tabular}
  \caption{Layers of the medium-sized CNN model
    \protect\inlinesh{fashion_mnist_medium.h5} for Fashion-MNIST}
  \label{tab:cnn-fashion-mnist-medium}
\end{table}

\subsection{Achieving Structural Criteria}

Structural coverage criteria such as NC or SSC focus on the patterns that appear in the outputs of ReLU activation functions.
Among the layers of @fashion_mnist_medium.h5@ (\cf \tablename~\ref{tab:cnn-fashion-mnist-medium}), this corresponds to the outputs of @activation@, @activation_1@, and @activation_2@.

\paragraph{Neuron Coverage with Pixel\-wise Optimisation (\(L_0\) Norm)}
As we are dealing with an image dataset, let us first exercise the search engine that operates via pixel\-wise optimisation.
This essentially amounts to an exploration of \(L_0\) norm balls, in search for a new image that satisfies the test target and that differs from the candidate image in as few pixels (or colour channels) as possible.

In this case, we explicitly select ReLU activation layers (as arguments to the @--layers@ flag), ask to save every generated input image into the output directory @outs/fm-medium/nc-l0@ by giving the @--save-all-tests@ flag, and restrict ourselves to \num{100} iterations so as to obtain reasonable running times for the whole command (under a minute on a 3,4GHz Quad Core Intel i7 CPU with 16GB of memory):
% !deepconcolic nc fashion-mnist medium l0
\begin{cmds}
python3 -m deepconcolic.main --outputs outs/fm-medium/nc-l0 --dataset fashion_mnist --model saved_models/fashion_mnist_medium.h5 --layers activation activation_1 activation_2 --criterion nc --norm l0 --save-all-tests --max-iterations 100
\end{cmds}
\begin{textcode}
…
Randomly selecting an input from test data.
\end{textcode}
When given the above arguments, \DeepConcolic first constructs an initial singleton test set \(\Xtest_0\) by randomly selecting an input that is correctly classified by the model.
This seed input is taken from the test data \Xtest (provided as part of the Fashion-MNIST dataset).
Then the initial neuron coverage is shown:
\begin{textcode}
…
#0 NC: 36.49130629%
\end{textcode}
and the search engine subsequently alters the seed and newly generated inputs in order to increase the overall amount of neurons activated by the set of test cases under construction:
\begin{textcode}
| Targeting activation of (1, 4, 27) in conv2d
#1 NC: 36.95942934% with new test case at L0 distance 1: passed
| Targeting activation of (1, 19, 27) in conv2d
#2 NC: 37.40897607% with new test case at L0 distance 1: passed
…
| Targeting activation of (23, 4, 12) in conv2d
#63 NC: 50.53499777% with new test case at L0 distance 1: adversarial
| Targeting activation of (15, 21, 7) in conv2d
#64 NC: 50.53499777% after failed attempt
…
| Targeting activation of (19, 11, 15) in conv2d
#100 NC: 56.94010997% with new test case at L0 distance 1: passed
Terminating after 100 iterations: 92 tests generated, 4 of which are adversarial.
\end{textcode}
The last line output to the console shows a short summary of the execution.
Every generated test image is saved within the directory @outs/fm-medium/nc-l0@.
In the case of the command at hand, most of them consist in versions of the seed image where the brightness of individual pixels have been altered.
As an illustration, we show below a generated input on the left, the original candidate input in the middle, and the diff on the right; in the latter a red pixel indicates a change of value from the original to the generated input.

\begin{center}
  \includegraphics[width=6cm]{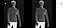}%
\end{center}

\paragraph{Neuron Coverage with \(L_∞\) Norm and a Concolic Search Engine}
Turning to a concolic search engine, we can employ a similar command line and switch to a norm that allows more global perturbations to input images.
% !deepconcolic nc fashion-mnist medium linf
\begin{cmds}
python3 -m deepconcolic.main --outputs outs/fm-medium/nc-linf --dataset fashion_mnist --model saved_models/fashion_mnist_medium.h5 --layers activation --criterion nc --norm linf --save-all-tests --max-iterations 10
\end{cmds}
Note that we are only considering the first activation layer (@activation@).
Indeed, the number of linear variables and constraints grows linearly with the depths of considered layers.
Thus, restricting ourselves to increase neuron coverage at the shallowest layer allows us to obtain reasonable overall execution times with the LP solver that ships with @pulp@.
Other activation layers of this model can additionally be considered when a more advanced LP solver is used, like CPLEX or GUROBI (see Section~\ref{sec:check-inst-solv}).

Moreover, the computational complexity of the search increases significantly \wrt the pixel-wise optimisation above: we therefore restrict ourselves to observing the first 10 iterations of \DeepConcolic so the command terminates within reasonable time (\red{less than 15 minutes}).

\begin{textcode}
…
Randomly selecting an input from test data.
#0 NC: 37.49075444%
| Targeting activation of (8, 11, 22) in conv2d
#1 NC: 37.50924556% with new test case at Linf distance 0.015686304195254464: passed
| Targeting activation of (7, 7, 29) in conv2d
#2 NC: 37.55085059% with new test case at Linf distance 0.04705884900747559: passed
…
#10 NC: 37.83746302% with new test case at Linf distance 0.09019610390943639: passed
Terminating after 10 iterations: 10 tests generated, 0 of which is adversarial.
\end{textcode}
The result of this command is similar to the one above.
In this case, however, the reported distances indicate the maximal change of value of any pixel instead of the number of changed pixels.

\begin{center}
  \includegraphics[width=6cm]{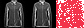}%
\end{center}

\paragraph{Sign-Sign Coverage with Concolic Search Engine}
Other coverage metrics and associated search engines can also be used, such as the MC/DC-style criteria.
Note the layers given to @--layers@ are \emph{decision} layers only, therefore we omit the first layer (@activation@) below, which can only be considered as encoding a set of conditions in MC/DC criteria.
We omit the deepest layer (@activation_2@) as well for the same reason as above.
In this case, however, the computational complexity of the symbolic search further increases \wrt the basic neuron coverage considered above, with typical LP problems of more than \num{100 000} constraints for the command below.
Still, a reasonable termination time of less than 15 minutes can be obtained when using advanced LP solvers (CPLEX in this instance).
% !deepconcolic ssc ssclp fashion-mnist medium linf large-lps
\begin{cmds}
python3 -m deepconcolic.main --outputs outs/fm-medium/ssclp-linf --dataset fashion_mnist --model saved_models/fashion_mnist_medium.h5 --layers activation_1 --criterion ssclp --norm linf --init 100 --save-all-tests --max-iterations 10
\end{cmds}
\begin{textcode}
…
Initializing with 100 randomly selected test cases that are correctly classified.
#0 SSC: 0.00000000%
| Targeting decision 82 in dense, subject to condition (5, 5, 26) in conv2d_1
#1 SSC: 0.00086207% with new test case at Linf distance 0.031372578471314694: passed
| Targeting decision (2, 4, 0) in conv2d_1, subject to condition (0, 0, 0) in activation
#2 SSC: 0.00086207% after failed attempt
…
#10 SSC: 0.00344828% with new test case at Linf distance 0.058823529411764774: passed
Terminating after 10 iterations: 3 tests generated, 0 of which is adversarial.
\end{textcode}

% Albeit relating to the combined behaviours of pairs of adjacent
% layers,
The patterns of changed pixels for generated tests often exhibit similar structures as the ones obtained above for neuron coverage:
\begin{center}
  \includegraphics[width=6cm]{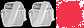}%
\end{center}
We additionally observe that the reported coverage increases very slowly with each successful iteration.
This stems from the definition of sign-sign coverage that induces a combinatorial enumeration of all possible sets of condition patterns for each decision neuron.
Also note the reported coverage starts at 0\% even though \num{100} test cases have been used to construct \(X_0\); this stems from the combinatorial nature of the coverage criterion, for which an accurate accounting based on all (unordered) elements of \(X_0\) would either be too computationally expensive, or lack a precise meaning.
% the purpose of efficiency, as an accurate computation of this initial
% coverage for large that \DeepConcolic does not account for this set
% for initial coverage due to the combinatorial nature of the criterion.

\paragraph{Sign-Sign Coverage with GA-based Search Engine}
The GA-based engine for sign-sign coverage (enabled with %
@--criterion ssc@) relaxes the aforementioned limitation by allowing users to specify a ratio of condition neurons whose activations need to be altered independently (instead of always a single one for the strict criterion as above); this ratio is specified with @--mcdc-cond-ratio@.
The command below targets the same sign-sign coverage criterion as above by using such an engine.
Due to the relaxation in the definition of the criterion, this generation approach reports higher coverage measures and is usually more efficient than its concolic counterpart (it takes about 12 minutes to terminate).
% !deepconcolic ssc fashion-mnist medium linf
\begin{cmds}
python3 -m deepconcolic.main --outputs outs/fm-medium/ssc-linf --dataset fashion_mnist --model saved_models/fashion_mnist_medium.h5 --layers activation_1 activation_2 --criterion ssc --norm linf --init 100 --mcdc-cond-ratio 0.1 --save-all-tests --max-iterations 10
\end{cmds}
\begin{textcode}
…
Initializing with 100 randomly selected test cases that are correctly classified.
#0 SSC: 0.00000000%
| Targeting decision (1, 4, 43) in conv2d_1, subject to any condition
#1 SSC: 0.01077586% with new test case at Linf distance 0.007843165304146527: passed
| Targeting decision 80 in dense, subject to any condition
#2 SSC: 0.45215517% with new test case at Linf distance 0.007843166706608784: passed
…
#10 SSC: 1.36724138% with new test case at Linf distance 0.007843166940352475: passed
Terminating after 10 iterations: 7 tests generated, 0 of which is adversarial.
\end{textcode}

\begin{center}
  \includegraphics[width=6cm]{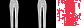}%
\end{center}

\subsection{Achieving High-level Criteria}

Let us now turn to the higher-level criteria available in \DeepConcolic.
As mentioned in Section~\ref{sec:high-level-criteria}, such criteria are defined based on a combination of a dimensionality reduction technique, and the construction of a Bayesian Network that allows one to capture how a given test dataset exercises high-level features that have been learned by hidden layers of the DNN.

In principle the high-level (BN-based) criteria can be used to investigate how a test dataset exercises a set of \LFs that has been learned from the training dataset and internally represented by \emph{any layer} of the CNNs% @fashion_mnist_medium.h5@
.
For the purposes of this tutorial, however, we will restrict ourselves to \LFs learned by @fashion_mnist_medium.h5@ at layers @activation@, @activation_1@, and @activation_2@ as well.

\paragraph{Extraction of Hidden Features via Dimensionality Reduction}

The following command performs dimensionality reduction on the neuron values of the activation layers based on \num{10 000} samples of training data (that ships with the Fashion-MNIST dataset).
For this example, we have opted to concentrate on the 3 principal components for each layer, that are partitioned into 5 intervals each to obtain a BN of 9 nodes and 18 edges.
The tool constructs the structure of the BN abstraction (this should take less than a minute), and saves it into @outs/fm-medium/bn.pkl@.
% !deepconcolic dbnc create fashion-mnist medium pca
\begin{cmds}
python3 -m deepconcolic.dbnabstr --dataset fashion_mnist --model saved_models/fashion_mnist_medium.h5 create outs/fm-medium/bn.pkl --feature-extraction pca --num-features 3 --num-intervals 3 --extended-discr --layers activation activation_1 activation_2 --train-size 10000
\end{cmds}
\begin{textcode}
…
Using extended 3-bin discretizer with uniform strategy for layer activation
Using extended 3-bin discretizer with uniform strategy for layer activation_1
Using extended 3-bin discretizer with uniform strategy for layer activation_2

| Given 10000 classified training sample
| Extracting and discretizing features for layer activation...
| Extracted 3 features
| Discretization of feature 0 involves 5 intervals
| Discretization of feature 1 involves 5 intervals
| Discretization of feature 2 involves 5 intervals
| Discretized 3 features
| Extracting and discretizing features for layer activation_1...
| Extracted 3 features
| Discretization of feature 0 involves 5 intervals
| Discretization of feature 1 involves 5 intervals
| Discretization of feature 2 involves 5 intervals
| Discretized 3 features
| Extracting and discretizing features for layer activation_2...
| Extracted 3 features
| Discretization of feature 0 involves 5 intervals
| Discretization of feature 1 involves 5 intervals
| Discretization of feature 2 involves 5 intervals
| Discretized 3 features
| Captured variance ratio for layer activation is 16.60%
| Captured variance ratio for layer activation_1 is 18.60%
| Captured variance ratio for layer activation_2 is 46.70%
| Created Bayesian Network of 9 nodes and 18 edges.
Dumping abstraction into `outs/fm-medium/bn.pkl'... done
\end{textcode}

When using PCA as a feature extraction technique, the output to the console reports, for each layer, the total amount of variance in its neuron values that is captured by the extracted principal components.
Here, we can observe that using 3 features for the first 2 layers captures less than 20\% of their respective variance: this basically means that many changes of neuron values for these layers are not reflected in significant-enough changes to any of the 3 first principal components.
Further tuning of the parameters above could be employed to improve these figures.
Yet, for the sole purposes of test case generation, even a small amount of captured variance in hidden layers is enough to obtain new inputs that exhibit learned features.
We will therefore reuse the computed abstraction below.

The BN abstraction as computed above allows us to define high-level coverage criteria that give an account on how extracted \LFs are exercised by a given test set.
To define such coverage metrics and criteria, the idea is to identify
% concolic search engine tries to synthesise new inputs that expand
ranges of values obtained for \LFs that are deemed not exercised enough: this is done internally via a partitioning of the \LFs into a set of intervals, and the marginal probabilities in the BN give a measure of the amount of test cases that exercises each \LFI.

\paragraph{Bayesian-network Feature Coverage Criterion}
% The BN abstraction as computed above allows us to define a high-level
% coverage criterion that gives an account on how extracted \LFs are
% exercised by a given test set.
The first coverage metric that we define based on the BN simply reports the ratio of intervals that are exercised.
To increase this coverage, the concolic search engine tries to synthesise new inputs that expand the range of values obtained for a \LF that is deemed not exercised enough.
The following command defines a feature coverage criterion based on the abstraction computed above, and then performs 10 iterations of concolic test case generation based on an initial set of 100 test cases (it should terminate within 10 minutes).
% !deepconcolic dbnc fashion-mnist medium bfc linf
\begin{cmds}
python3 -m deepconcolic.main --outputs outs/fm-medium/bfc-linf --dataset fashion_mnist --model saved_models/fashion_mnist_medium.h5 --bn-abstr outs/fm-medium/bn.pkl --criterion bfc --norm linf --save-all-tests --init 100 --max-iterations 10
\end{cmds}
\begin{textcode}
…
Initializing with 100 randomly selected test cases that are correctly classified.
#0 BFC: 86.66666667%
| Targeting interval (-inf, -66.3] of feature 0 in layer activation (from test 49)
#1 BFC: 86.66666667% with new test case at Linf distance 0.047058849942450465: passed
| Targeting interval (-inf, -66.3] of feature 0 in layer activation (from test 100)
#2 BFC: 86.66666667% with new test case at Linf distance 0.07058823529411773: passed
…
| Targeting interval (-inf, -66.3] of feature 0 in layer activation (from test 56)
#9 BFC: 86.66666667% after failed attempt
| Targeting interval (-inf, -66.3] of feature 0 in layer activation (from test 7)
#10 BFC: 86.66666667% after failed attempt
Terminating after 10 iterations: 5 tests generated, 0 of which is adversarial.
\end{textcode}

First of all, observe that the reported coverage does not always increase with each successful generation of a test case.
The reason behind this behaviour of \DeepConcolic is twofold: this is in part due to the discrete nature of our definitions for high-level coverage metrics, that only count some amount of test cases that fall within given \LFIs.
On the other hand, both the linear dimensionality reduction technique and the LP encoding performed by the search engine introduce over-approximations that lead to inputs not meeting their intended test targets.

Still, by examining the changes of pixel colours, we observe that generated inputs tend to exhibit alterations that show how some semantics of the original input is captured and exercised to produce new images.
\begin{center}
  \includegraphics[width=6cm]{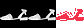}%
\end{center}
Further, we find empirically that these new inputs tend to be closer to reaching their test target than their respective candidate inputs.
\DeepConcolic therefore retains them as part of the set of legitimate generated test cases.

\paragraph{Bayesian-network Feature Coverage Criterion with
  Pixel\-wise Optimisation}
\DeepConcolic allows one to exercise the pixel\-wise optimisation search engine for increasing high-level coverage.
% Similarly to the case above, w
We can thus combine the BFC target as above with \(L_0\) norm exploration:
% !deepconcolic dbnc fashion-mnist medium bfc l0
\begin{cmds}
python3 -m deepconcolic.main --outputs outs/fm-medium/bfc-l0 --dataset fashion_mnist --model saved_models/fashion_mnist_medium.h5 --bn-abstr outs/fm-medium/bn.pkl --criterion bfc --norm l0 --save-all-tests --init 100 --max-iterations 10
\end{cmds}
% rng seed: 2459485544
% ≈ 13s
\begin{textcode}
…
Starting tests for criterion BFC with norm L0 (10 max iterations).
Reporting into: outs/fm-medium/bfc-l0/BFC_L0_report.txt
Initializing with 100 randomly selected test cases that are correctly classified.
#0 BFC: 86.66666667%
| Targeting interval (-inf, -66.3) of feature 0 in layer activation (from test 83)
#1 BFC: 88.88888889% with new test case at L0 distance 5: passed
| Targeting interval (-inf, -54.5) of feature 2 in layer activation (from test 45)
#2 BFC: 91.11111111% with new test case at L0 distance 5: passed
| Targeting interval [106, inf) of feature 1 in layer activation (from test 31)
#3 BFC: 91.11111111% after failed attempt
| Targeting interval [106, inf) of feature 1 in layer activation (from test 30)
#4 BFC: 91.11111111% after failed attempt
| Targeting interval [106, inf) of feature 1 in layer activation (from test 65)
#5 BFC: 91.11111111% after failed attempt
| Targeting interval (-inf, -89.6) of feature 1 in layer activation (from test 8)
#6 BFC: 93.33333333% with new test case at L0 distance 14: passed
| Targeting interval [123, inf) of feature 2 in layer activation (from test 102)
#7 BFC: 95.55555556% with new test case at L0 distance 5: adversarial
| Targeting interval [106, inf) of feature 0 in layer activation (from test 32)
#8 BFC: 95.55555556% after failed attempt
| Targeting interval [106, inf) of feature 1 in layer activation (from test 35)
#9 BFC: 95.55555556% after failed attempt
| Targeting interval [106, inf) of feature 0 in layer activation (from test 3)
#10 BFC: 95.55555556% after failed attempt
Terminating after 10 iterations: 4 tests generated, 1 of which is adversarial.
\end{textcode}
% python3 -m deepconcolic.main --outputs /tmp/outs/fm-medium/bfc-l0 --dataset fashion_mnist --model saved_models/fashion_mnist_medium.h5 --bn-abstr outs/fm-medium/bn.pkl --criterion bfc --norm l0 --save-all-tests --init 100 --max-iterations 1000

\begin{minipage}[t]{.48\textwidth}
  \centering
  Correctly classified inputs:
  \includegraphics[width=6cm]{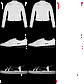}
\end{minipage}
\hfill%
\begin{minipage}[t]{.48\textwidth}
  \centering
  Adversarial example:
  \includegraphics[width=6cm]{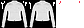}
\end{minipage}

\paragraph{Bayesian-network Feature-dependence Coverage Criterion}

The feature-dependence coverage criterion is the BN-based counterpart of MC/DC-style structural criteria like sign-sign above, as it only targets combinations of \LFIs in successive layers that are not exercised in the test dataset so far.
This command should terminate within 1 hour when an advanced LP solver with appropriate python bindings is used (\eg CPLEX---\cf Section~\ref{sec:check-inst-solv}).
% !deepconcolic dbnc fashion-mnist medium bfdc linf
\begin{cmds}
python3 -m deepconcolic.main --outputs outs/fm-medium/bfdc-linf --dataset fashion_mnist --model saved_models/fashion_mnist_medium.h5 --bn-abstr outs/fm-medium/bn.pkl --criterion bfdc --norm linf --save-all-tests --init 100 --max-iterations 10
\end{cmds}
\begin{textcode}
…
Initializing with 100 randomly selected test cases that are correctly classified.
#0 BFdC: 81.39733333%
| Targeting interval [-41.5, -10.1] of feature 2 in layer activation_1, subject to feature intervals (1, 1, 1) in layer activation (from test 55)
#1 BFdC: 81.39733333% after failed attempt
…
#4 BFdC: 81.39733333% after failed attempt
| Targeting interval [-5.23, 25.5] of feature 1 in layer activation_1, subject to feature intervals (1, 3, 2) in layer activation (from test 1)
#5 BFdC: 86.20444444% with new test case at Linf distance 0.08235297086192112: passed
…
#10 BFdC: 86.66666667% after failed attempt
Terminating after 10 iterations: 2 tests generated, 0 of which is adversarial.
\end{textcode}

Regarding generated inputs, we observe that focusing on a combinatorial coverage does not fundamentally change the overall nature of generated images; this behaviour is in line with what we observed in the case of structural coverage above.
\begin{center}
  \includegraphics[width=6cm]{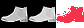}%
\end{center}

\subsection{Naive Search Engine: Fuzzing}
\DeepConcolic supports an experimental fuzzing engine, that can be used with the command-line flag \shparambox{--fuzzing}.
Fuzzing follows a simple approach to testing, by randomly mutating well-formed inputs and then running the neural network with those mutated inputs in the hope of triggering adversarial examples.
Although naive, this approach has been remarkably efficient in finding bugs in programs that had never been fuzzed% \nb{Citation?}
.
While the ultimate goal is to interleave the fuzzing engine and the symbolic engine in \DeepConcolic for more effectively exploring the vulnerabilities in neural networks, this is not implemented yet.

Still, the engine can be run via similar command lines.
For instance, the following command will fuzz the Fashion-MNIST model @fashion_mnist_large.h5@ with the seed input images contained in the directory specified by means of the argument to \shparambox{--inputs}.
% !deepconcolic fuzzing fashion-mnist large
\begin{cmds}
python3 -m deepconcolic.main --fuzzing --model saved_models/fashion_mnist_large.h5 --num-processes 2 --inputs data/mnist-seeds/ --outputs outs/fm-large/fuzzing --input-rows 28 --input-cols 28 --num-tests 10
\end{cmds}
This should result in:
\begin{itemize*}[(i)]
\item one @mutants@ folder;
\item one @advs@ folder for adversarial examples; and
\item a file @adv.list@ that lists the commands for validating the adversarial examples, and can be used to retrieve a list of all adversarial examples found.
\end{itemize*}
The @--num-processes 2@ means that a parallelism of two processes will be used to perform the fuzzing.
By default, fuzzing will terminate after \num{1000} iterations---this can be configured by using the \shparambox{--num-tests} command-line argument.
A configuration @--num-tests 1000@ should be interpreted in the way that $n\times 1000$ test cases would be generated if a number of $n$ threads are used in the fuzzing.
Note the example above runs for about three hours (on an Intel Core i7-7820HQ, 16GB Memory)% \todo{NB: dumb question: Why does it need to sleep 4 seconds after each iteration? \\ YS: The 4 seconds are for sub-threads to run; after 4 seconds, they will be forced to terminate (if their executions have not terminated. It would be better if the this ``4 seconds'' is configurable)}
; of course, this execution time varies linearly with the argument given to @--num-tests@.
Also note that the fuzzing engine has only ever been tested on an Ubuntu system, and it is very likely that it may not run properly on non-GNU/Linux-based systems like Windows.

\section{Example: Human Activity Recognition% (Preprocessed)
}
\label{sec:human-activ-recogn}

% \begin{redenv}The second case study that we selected considers the classification of
% human activities from sensor data~\citep{anguita2013public-HAR}.
% Instead of 2-dimensional images as above, inputs consist in
% 1-dimensional arrays of 561 features in total, that comprise
% preprocessed (\eg normalised) acceleration, velocity, gyroscopic data
% from sensors\footnote{This dataset is originally from
%   \url{https://archive.ics.uci.edu/ml/datasets/human+activity+recognition+using+smartphones};
%   it can be found in preprocessed form at
%   \url{https://www.openml.org/d/1478}.}.
% The classification task at hand is to determine whether each sample
% corresponds to sensor readings obtained while the subject performed
% one of six activities: walking, walking upstairs, walking downstairs,
% sitting, standing, or laying.

% This particular task allows us to concentrate on \DeepConcolic's
% abilities to handle non-image data (\ie in our case we are dealing
% with inputs that consist of 1-dimensional arrays of normalised
% input features).
% \end{redenv}

Let us now turn to the HAR classification task presented in Section~\ref{sec:uci-har}.

\subsection{Secific Challenges}
\label{sec:secific-challenges}

% \paragraph{Dealing with Well-formed\-ness Constraints for Inputs}
As far as \DeepConcolic is concerned, the additional challenge in dealing with the HAR dataset stems from the need to rely on additional intrinsic constraints on input features to rule out implausible inputs.
For this, we make use of a post-filter.
Indeed, determining a suitable distance measure to compare inputs and determine the plausibility of generated ones depends on the semantics of each individual feature.
Such intrinsic constraints do not arise in the case of images, since one can consider that any combination of colour for every pixel constitutes a valid image.

% \red{As we additionally want to enforce some further constraints on
%   any generated input \(x'\) \wrt the input \(x\) from which it is
%   derived}, we define a post-filter based on a \emph{Local Outlier
%   Factor}~\citep{Breunig2000LOF}.
% This basically consists in constructing a tree-like structure from (a
% subset of the training data \Xtrain), that allows to efficiently
% estimate a distance between any new input and elements \Xtrain: the
% idea is then to filter out any new input that is too far from the
% training dataset \wrt the distance considered.
% The underlying distance can be customised based on the desired measure
% of similarity; in our case, we chose the cosine distance, so that
% inputs are kept if they show enough correlation with an element from
% \Xtrain.

\paragraph{Investigating Generated Inputs}

\begin{figure}
  \centering%
  \input{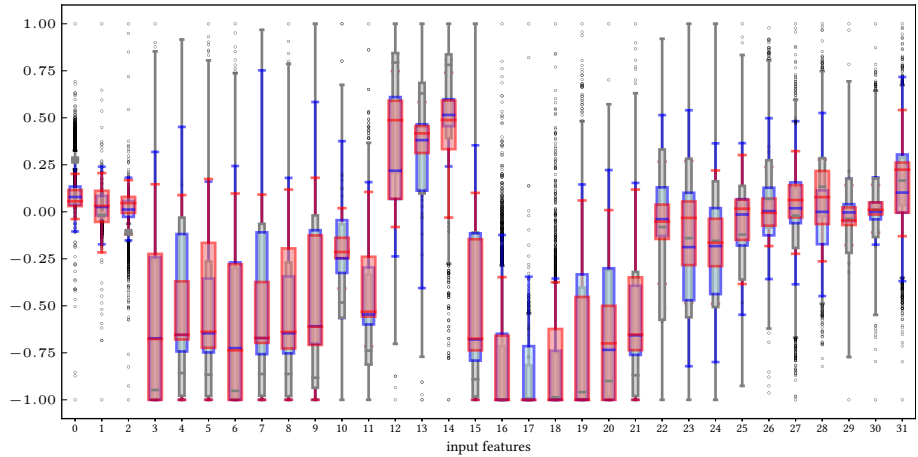}%
  \caption{Distribution of 32 input feature values for the HAR dataset (out of 561): grey bars represent the distribution of all values from the the training dataset, blue bars show the same distributions for an example set of \num{1714} inputs generated by \DeepConcolic, that are correctly classified by \Nhs; similarly, red bars show the values for \num{54} generated inputs that are incorrectly classified (\ie adversarial examples).}
  \label{fig:har-input-distributions}
\end{figure}
Since the inputs do not directly consist in images that can easily be inspected visually, we use simple plots as in \figurename~\ref{fig:har-input-distributions} to represent the distribution of values for a subset of all input features.
We generate these plots by using a HAR-specific @utils.harviz@ companion script included in \DeepConcolic's source code:
% !silent har plot
\begin{cmds}
python3 -m utils.harviz -h
\end{cmds}

% Designers of the DNN, however, may be able to use such a plot to
% conduct a preliminary visual assessments that generated inputs .
Expert designers with additional knowledge about the kind of sensor input data at hand would surely compute more advanced evaluation procedures for assessing whether the generated inputs satisfy the aforementioned plausibility criterion.

\subsection{Training or Downloading Models}

Similarly to the case of Fashion-MNIST, \DeepConcolic's source code includes a script that enables one to construct and train a dense DNN for this dataset.
It saves models under @/tmp@, and can be executed from within the the repository's root directory via
% For instance, we provide a script at train a DNN model for this
% dataset with
the following command (this script automatically retrieves the dataset from the Internet):
% !train deepconcolic har dense
\begin{cmds}
python3 -m deepconcolic.gen_har
\end{cmds}
Generated models can then be copied from @/tmp@ into the @saved_models@ directory (to be created, if necessary).
Our server also hosts a pre-trained model for this dataset, whose architecture is described in \tablename~\ref{tab:dnn-har-dense}:
% !silent download deepconcolic har dense
\begin{cmds}
wget -P saved_models https://cgi.csc.liv.ac.uk/~acps/models/har_dense.h5
\end{cmds}
\begin{table}%
  \centering\smaller
  \begin{tabular}{clrr}\hline%
    Layer    & Name \& Function specification   & Output shape & \#parameters  \\ \hline
    \layer 0 & \texttt{dense} (dense + ReLU)    & 192          & \num{107 904} \\
    \layer 1 & \texttt{dense\_1} (dense + ReLU) & 128          & \num{24 704}  \\
    \layer 2 & \texttt{dropout} (dropout)       & 128          & 0             \\
    \layer 3 & \texttt{dense\_2} (dense + ReLU) & 92           & \num{11 868}  \\
    \layer 4 & \texttt{dense\_3} (dense + ReLU) & 64           & \num{5 952}   \\
    \layer 5 & \texttt{dense\_4} (dense)        & 6            & \num{390}     \\
    \layer 6 & \texttt{activation} (softmax)    & 6            & 0             \\ \hline
  \end{tabular}%
  \caption{Layers of the dense DNN model
    \protect\inlinesh{har_dense.h5} for the HAR dataset}%
  \label{tab:dnn-har-dense}%
\end{table}%

\subsection{Achieving Structural Criteria}

In principle, \DeepConcolic's structural criteria and associated search engines may also be used for testing the model \Nhs that we have designed and trained for the HAR classification task.

% \paragraph{Neuron Coverage with Feature\-wise Optimisation (\(L_0\) Norm)}
For instance, with the command below \DeepConcolic selects the ``pixel\-wise'' optimisation search engine, that attempts to find new inputs that increase neuron coverage by altering as few input feature values as possible from the given candidate inputs.
% !deepconcolic nc har dense l0
\begin{cmds}
python3 -m deepconcolic.main --outputs outs/har-dense/nc-l0 --dataset OpenML:har --model saved_models/har_dense.h5 --layers dense dense_{1,2,3} activation --criterion nc --norm l0 --save-all-tests --max-iterations 100
\end{cmds}
\begin{textcode}
…
Starting tests for criterion NC with norm L0 (100 max iterations).
Reporting into: outs/har-dense/nc-l0/NC_L0_report.txt
Randomly selecting an input from test data.
#0 NC: 99.37759336%
| Targeting activation of 2 in dense_4
#1 NC: 99.58506224% with new test case at L0 distance 4: passed
| Targeting activation of 3 in dense_4
#2 NC: 99.79253112% with new test case at L0 distance 10: passed
| Targeting activation of 5 in dense_4
#3 NC: 100.00000000% with new test case at L0 distance 1: passed
Terminating after 3 iterations: 3 tests generated, 0 of which is adversarial.
\end{textcode}

% \paragraph{Neuron Coverage with \(L_∞\) Norm and a Concolic Search Engine}
A concolic search engine is also selected when exploring the \(L_∞\) norm ball:
% !deepconcolic nc har dense linf
\begin{cmds}
python3 -m deepconcolic.main --outputs outs/har-dense/nc-linf --dataset OpenML:har --model saved_models/har_dense.h5 --layers dense dense_{1,2,3} activation --criterion nc --norm linf --save-all-tests --max-iterations 100
\end{cmds}
\begin{textcode}
…
Starting tests for criterion NC with norm Linf (100 max iterations).
Reporting into: outs/har-dense/nc-linf/NC_Linf_report.txt
Randomly selecting an input from test data.
#0 NC: 99.37759336%
| Targeting activation of 2 in dense_4
Infeasible
#1 NC: 99.37759336% after failed attempt
| Targeting activation of 3 in dense_4
Infeasible
#2 NC: 99.37759336% after failed attempt
| Targeting activation of 5 in dense_4
Infeasible
…
#99 NC: 99.37759336% after failed attempt
| Targeting activation of 0 in dense_4
Infeasible
#100 NC: 99.37759336% after failed attempt
Terminating after 100 iterations: 0 test generated, 0 of which is adversarial.
\end{textcode}
In both cases above, however, we observe that a structural coverage that only takes activation patterns into account is not appropriate to test \Nhs, as even a single input taken randomly is enough to achieve near-100\% neuron coverage.

In fact, this stems from a technicality in the architecture of \Nhs, where all considered layers integrate ReLU activation functions (except @dense_4@, that does not integrate an activation function as it is directly followed by a softmax layer---see \tablename~\ref{tab:dnn-har-dense}).
Since this eliminates every negative value most layers, every activation pattern observed by \DeepConcolic includes activated neurons only.
% \red{From this, we can guess that the DNN has learned based on

Furthermore, the LP-based concolic search engine is unable to derive a concrete input that triggers the activation of neuron 0 in layer @dense_4@ from the only seed input in \(X_0\).
Indeed, we can actually see from the trace above that every generated LP problem is infeasible.
This means that no test exists that triggers the sought-after activation within the \(L_∞\) norm ball around the seed input (and that meets the input bound constraints induced by \dminHard and \dminNoise)

\subsection{Achieving High-level Criteria}

Let us first use \DeepConcolic's companion tool dedicated to the creation of BN-based abstractions, and focus on the output of every activation layer.
The computation of the abstraction is done with the \shparambox{create} sub-command of this tool:
% !deepconcolic dbnc create har dense pca
\begin{cmds}
python3 -m deepconcolic.dbnabstr --dataset OpenML:har --model saved_models/har_dense.h5 create outs/har-dense/bn.pkl --feature-extraction pca --num-features 3 --num-intervals 3 --extended-discr --layers dense dense_{1,2,3} activation
\end{cmds}
\begin{textcode}
…
| Given 7724 classified training sample
| Extracting features for layer dense...
| Extracted 3 features
| Discretizing features for layer dense...
| Discretization of feature 0 involves 3 intervals
| Discretization of feature 1 involves 3 intervals
| Discretization of feature 2 involves 3 intervals
| Discretized 3 features
| Extracting features for layer dense_1...
| Extracted 3 features
| Discretizing features for layer dense_1...
| Discretization of feature 0 involves 3 intervals
| Discretization of feature 1 involves 3 intervals
| Discretization of feature 2 involves 3 intervals
| Discretized 3 features
| Extracting features for layer dense_2...
| Extracted 3 features
| Discretizing features for layer dense_2...
| Discretization of feature 0 involves 3 intervals
| Discretization of feature 1 involves 3 intervals
| Discretization of feature 2 involves 3 intervals
| Discretized 3 features
| Extracting features for layer dense_3...
| Extracted 3 features
| Discretizing features for layer dense_3...
| Discretization of feature 0 involves 3 intervals
| Discretization of feature 1 involves 3 intervals
| Discretization of feature 2 involves 3 intervals
| Discretized 3 features
| Extracting features for layer activation...
| Extracted 3 features
| Discretizing features for layer activation...
| Discretization of feature 0 involves 3 intervals
| Discretization of feature 1 involves 3 intervals
| Discretization of feature 2 involves 3 intervals
| Discretized 3 features
| Captured variance ratio for layer dense is 64.74%
| Captured variance ratio for layer dense_1 is 66.94%
| Captured variance ratio for layer dense_2 is 75.86%
| Captured variance ratio for layer dense_3 is 76.04%
| Captured variance ratio for layer activation is 61.03%
| Created Bayesian Network of 15 nodes and 36 edges.
Dumping abstraction into `outs/har-dense/bn.pkl'... done
\end{textcode}
Regarding feature extraction, we can first observe that a substantial amount of all variance in the data can be captured with three \HFs at each layer considered.
The sub-command \shparambox{show} enables us to inspect the \LFIs for each layer and extracted feature:
% !deepconcolic dbnc show har dense
\begin{cmds}
python3 -m deepconcolic.dbnabstr --dataset OpenML:har --model saved_models/har_dense.h5 show outs/har-dense/bn.pkl
\end{cmds}
\begin{textcode}
…
Loading abstraction from `outs/har-dense/bn.pkl'... done
===  Extracted Features and Associated Intervals  ==============================
Layer       Feature    Intervals
----------  ---------  -----------------------------------------
dense       0          (-inf, -15.6), [-15.6, 155), [155, inf)
            1          (-inf, -3.13), [-3.13, 375), [375, inf)
            2          (-inf, -11.7), [-11.7, 29.1), [29.1, inf)
dense_1     0          (-inf, -9.06), [-9.06, 15.6), [15.6, inf)
            1          (-inf, -6.52), [-6.52, 19.1), [19.1, inf)
            2          (-inf, -11.2), [-11.2, 13.5), [13.5, inf)
dense_2     0          (-inf, -8.43), [-8.43, 15.8), [15.8, inf)
            1          (-inf, -6.71), [-6.71, 14.2), [14.2, inf)
            2          (-inf, -5.83), [-5.83, 10.5), [10.5, inf)
dense_3     0          (-inf, -7.16), [-7.16, 12.6), [12.6, inf)
            1          (-inf, -5.08), [-5.08, 11), [11, inf)
            2          (-inf, -7.05), [-7.05, 8.23), [8.23, inf)
activation  0          (-inf, -1.61), [-1.61, 1.81), [1.81, inf)
            1          (-inf, -1.53), [-1.53, 1.69), [1.69, inf)
            2          (-inf, -1.44), [-1.44, 1.78), [1.78, inf)
\end{textcode}
In this case as we have used an extended discretization strategy (@--extended-discr@) and three intervals, we can inspect the range of values covered by the training data for each extracted \HF by looking at the second interval in each row.

\paragraph{High-level Feature Coverage Criterion with Feature\-wise
  Optimisation}

When targeting BFC with the \(L_0\) norm for measuring distances between inputs for this model, \DeepConcolic uses the ``pixel\-wise'' optimisation search engine and attempts to find new inputs by altering as few values of input features as possible.
Observe that we do not specify any size for \(X_0\), which means that a single seed input is randomly drawn from the test samples that ship with the raw dataset.
% !deepconcolic dbnc har dense bfc l0 single-seed
\begin{cmds}
python3 -m deepconcolic.main --outputs outs/har-dense/bfc-l0-singleseed --dataset OpenML:har --model saved_models/har_dense.h5 --bn-abstr outs/har-dense/bn.pkl --criterion bfc --norm l0 --filters LOF --save-all-tests --max-iterations 1000
\end{cmds}
\begin{textcode}
…
Initializing LOF-based novelty estimator with 3000 training samples... done
LOF-based novelty offset is -1.5
Starting tests for criterion BFC with norm L0 (1000 max iterations).
…
Randomly selecting an input from test data.
#0 BFC: 33.33333333%
| Targeting interval [1.78, inf) of feature 2 in layer activation (from test 0)
#1 BFC: 33.33333333% after failed attempt
| Targeting interval [1.81, inf) of feature 0 in layer activation (from test 0)
#2 BFC: 33.33333333% after failed attempt
| Targeting interval (-inf, -1.53) of feature 1 in layer activation (from test 0)
#3 BFC: 33.33333333% after failed attempt
| Targeting interval (-inf, -5.08) of feature 1 in layer dense_3 (from test 0)
#4 BFC: 33.33333333% after failed attempt
| Targeting interval (-inf, -6.52) of feature 1 in layer dense_1 (from test 0)
#5 BFC: 33.33333333% with failed attempt at L0 distance 13: too far from original input
| Targeting interval [1.69, inf) of feature 1 in layer activation (from test 0)
#6 BFC: 33.33333333% after failed attempt
…
| Targeting interval [155, inf) of feature 0 in layer dense (from test 0)
#29 BFC: 33.33333333% after failed attempt
| Targeting interval [375, inf) of feature 1 in layer dense (from test 0)
#30 BFC: 33.33333333% after failed attempt
Unable to find a new candidate input!
Terminating after 30 iterations: 0 test generated, 0 of which is adversarial.
\end{textcode}

We get two useful pieces of information from the above output.
%
% \subparagraph{LOF-based Filter}
%
First, the @--filters LOF@ arguments triggers the initialisation of the LOF-based post-filter.
% estimator that relies on the computation of \emph{Local Outlier
%   Factors}~\citep{Breunig2000LOF}.
% When enabled, this device is used in addition to the oracle mechanism
% described previously, to detect and filter out generated inputs that
% can be considered outliers \wrt the raw training data.
In this particular instance, the estimator actually records a set \(X_{\mathtt{LOF}}\) of \num{3 000} inputs% using a specialised search
% tree that can be used to identify the \(k\)-nearest neighbours in
% \(X_{\mathtt{LOF}}\) that are the closest to any new input \(x\) \wrt
% some configurable distance metric
.
% The local outlier factor of \(x\) is then computed based on the
% density of its \(k\)-nearest neighbours, and this factor is used to
% determine the plausibility of \(x\) using a predefined threshold.
% that lies above some baseline threshold value used by the
% oracle to filter out implausible inputs.

% Of course, the choice of the distance metric is of paramount
% importance for this mechanism to be relevant and act as a useful
% oracle.
% By default the LOF-based filter uses the \emph{cosine distance} as a
% measure of distance between two inputs.
% This distance gives an appropriate indication of the relative
% similarity between two 1-dimensional data vectors, as it only
% considers the angle between the two normalised vectors and disregards
% the magnitude of each one of their respective features.
% % the respective features for 1-dimensional data vectors.

% \subparagraph{Influence of \(|X_0|\)}

We can also observe that the test generation algorithm stops prematurely as it is unable to find a suitable candidate input for any unmet test target at iteration @#30@.
In the case of high-level feature coverage, this basically means that no suitable candidate input can be identified for every \HFI that is left un\-exercised by the test dataset \(X\) (\ie the union of \(X_0\) and all generated inputs---none in the case above).

\paragraph{Increasing \(|X_0|\)}
Let us increase the size of \(X_0\) to help the test candidate selection heuristics:
% !deepconcolic dbnc har dense bfc l0 init100
\begin{cmds}
python3 -m deepconcolic.main --outputs outs/har-dense/bfc-l0-init100 --dataset OpenML:har --model saved_models/har_dense.h5 --bn-abstr outs/har-dense/bn.pkl --criterion bfc --norm l0 --filters LOF --save-all-tests --init 100 --max-iterations 1000
\end{cmds}
% time ≈ 35s
\begin{textcode}
…
Initializing with 100 randomly selected test cases that are correctly classified.
#0 BFC: 40.00000000%
| Targeting interval (-inf, -1.61) of feature 0 in layer activation (from test 1)
#1 BFC: 40.00000000% with new test case at L0 distance 1: passed
| Targeting interval (-inf, -1.61) of feature 0 in layer activation (from test 3)
#2 BFC: 40.00000000% with new test case at L0 distance 1: passed
…
| Targeting interval [375, inf) of feature 1 in layer dense (from test 100)
#998 BFC: 88.88888889% after failed attempt
| Targeting interval [375, inf) of feature 1 in layer dense (from test 120)
#999 BFC: 88.88888889% after failed attempt
| Targeting interval [375, inf) of feature 1 in layer dense (from test 119)
#1000 BFC: 88.88888889% after failed attempt
Terminating after 1000 iterations: 111 tests generated, 0 of which is adversarial.
\end{textcode}
This particular command outputs every generated test into a file @outs/har-dense/bfc-l0-init100/new_inputs.csv@.

This time, we can remark a substantial increase in the proportion of \HFIs that are exercised by the generated set of \num{111} tests.
We can further investigate the latter by using the @utils.harviz@ script:
% !deepconcolic dbnc har dense bfcplot l0 init100
\begin{cmds}
DC_MPL_BACKEND=pdf DC_MPL_FIG_RATIO=.5 python3 -m utils.harviz --outputs outs/har-dense/bfc-l0-init100/plot outs/har-dense/bfc-l0-init100 --features 32
\end{cmds}
\begin{figure}%
  \centering%
  \input{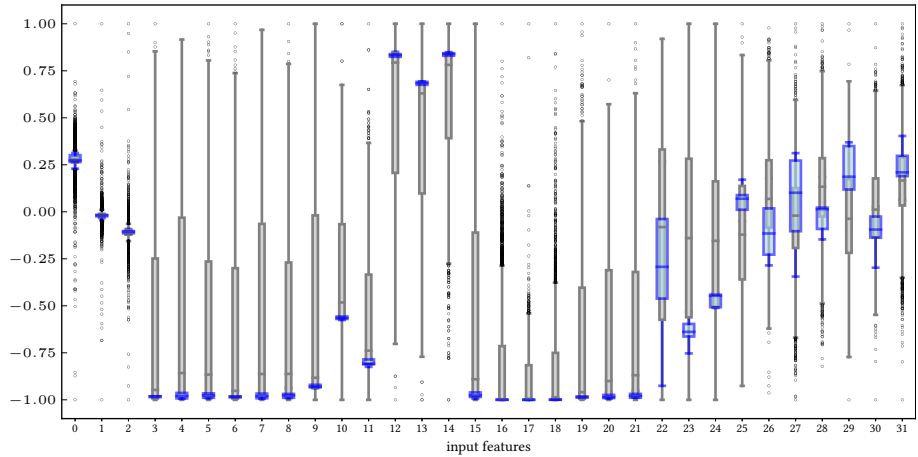}%
  \caption{Distribution of 32 selected input feature values for the HAR dataset (out of 561): grey bars represent the distribution of all values from the the training dataset, and blue bars indicate the distribution of feature values for the set of \num{111} inputs generated to achieve feature coverage with exploration of \(L_0\) norm ball.}%
  \label{fig:har-dense-l0-init100-distributions}%
\end{figure}%
The above command generates a file @outs/har-dense/bfc-l0-init100/plot/har-0.pdf@ which is as shown in \figurename~\ref{fig:har-dense-l0-init100-distributions}; substitute @DC_MPL_BACKEND=X11@ for @DC_MPL_BACKEND=pdf@ for an interactive plot.
Similarly to \figurename~\ref{fig:har-input-distributions}, this plot shows the distribution of 32 selected input feature values for the HAR dataset, where the blue bars indicate the distribution pertaining to the set of \num{111} generated inputs that attain 89\% high-level feature coverage (see above).
These inputs are:
\begin{enumerate*}[(i)]
\item correctly classified by @har_dense.h5@; and
\item close enough to original inputs in \(X_0\) \wrt the \(L_0\) norm (which means that only a few features of original inputs have been modified to form each new generated input).
\end{enumerate*}

\paragraph{High-level Feature Coverage Criterion with Concolic Search Engine}

Turning to a concolic search engine and the \(L_∞\) norm:
% !deepconcolic dbnc har dense bfc linf init100
\begin{cmds}
python3 -m deepconcolic.main --outputs outs/har-dense/bfc-linf-init100 --dataset OpenML:har --model saved_models/har_dense.h5 --bn-abstr outs/har-dense/bn.pkl --criterion bfc --norm linf --filters LOF --save-all-tests --init 100 --max-iterations 100
\end{cmds}
\begin{textcode}
Starting tests for criterion BFC with norm Linf (100 max iterations).
Reporting into: outs/har-dense/bfc-linf-init100/BFC_Linf_report.txt
Initializing with 100 randomly selected test cases that are correctly classified.
#0 BFC: 40.00000000%
| Targeting interval (-inf, -1.61) of feature 0 in layer activation (from test 3)
#1 BFC: 40.00000000% with new test case at Linf distance 0.0152358744: passed
| Targeting interval (-inf, -1.61) of feature 0 in layer activation (from test 15)
#2 BFC: 40.00000000% with new test case at Linf distance 0.03117379000000009: passed
| Targeting interval (-inf, -1.61) of feature 0 in layer activation (from test 22)
#3 BFC: 40.00000000% with new test case at Linf distance 0.04038073100000002: passed
| Targeting interval (-inf, -1.61) of feature 0 in layer activation (from test 31)
#4 BFC: 40.00000000% with failed attempt at Linf distance 0.21034023000000007: too far from original input
…
#98 BFC: 40.00000000% after failed attempt
| Targeting interval (-inf, -1.44) of feature 2 in layer activation (from test 12)
#99 BFC: 40.00000000% after failed attempt
| Targeting interval (-inf, -1.44) of feature 2 in layer activation (from test 46)
#100 BFC: 40.00000000% after failed attempt
Terminating after 100 iterations: 74 tests generated, 0 of which is adversarial.
\end{textcode}
\begin{figure}%
  \centering%
  \input{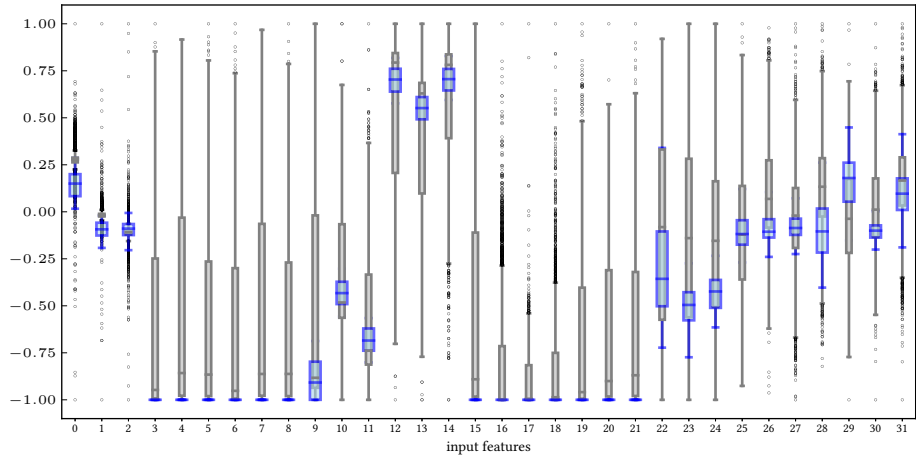}%
  \caption{Distribution of 32 selected input feature values for the HAR dataset (out of 561): grey bars represent the distribution of all values from the the training dataset, and blue bars indicate the distribution of feature values for the set of \num{74} inputs generated to achieve feature coverage with exploration of \(L_∞\) norm ball.}%
  \label{fig:har-dense-linf-init100-distributions}%
\end{figure}%
%

% % !deepconcolic dbnc har dense bfc linf init100 lbharder
% \begin{cmds}
% python3 -m deepconcolic.main --outputs outs/har-dense/bfc-linf-init100-lbharder --dataset OpenML:har --model saved_models/har_dense.h5 --bn-abstr outs/har-dense/bn.pkl --criterion bfc --norm linf --filters LOF --save-all-tests --init 100 --max-iterations 100 --lb-hard 5e-2 --lb-noise 1e-1
% \end{cmds}

% \begin{redenv}
% \subsection{Conclusions (so far)}
% \label{sec:conclusions-so-far}

% The case studies we have conducted indicate that structural coverage
% criteria such as NC and SSC are more efficient than high-level
% criteria at assessing the adversarial robustness of a neural network
% under consideration.
% These results seem to support the results of~\citet{LMXC2019}, which
% observed that testing techniques that achieve structural coverage
% criteria often perform similarly to techniques that purposefully
% search for adversarial examples.

% The fact that the concolic testing approach for achieving high-level
% criteria is able to generate new test inputs that exhibit more
% targeted changes of inputs feature than for structural counterparts,
% indicates that its effectiveness gears more towards interpret\-ability
% and data augmentation (as can be observed for the image classification
% case---and is yet to be confirmed but can be speculated for the HAR
% case study).
% \end{redenv}

\chapter{Testing Recurrent Neural Networks} %(written by Wei and Xiaowei)}
\label{part:testrnn}

Let us now turn to the \TestRNN tool for testing recurrent neural networks (RNNs).
This work specifically focuses on a widespread class of RNNs called long short-term memory models (LSTMs).
% \section{Introduction}
One particular challenge in handling such models lies in the internal temporal behaviour of the LSTM layers in processing sequential inputs.
\TestRNN implements new coverage metrics that we have designed to address this issue:
% We have designed new coverage metrics to consider the internal behaviour of the LSTM layers in processing sequential inputs.
we consider not only a tight quantification of temporal behaviours called \emph{Temporal Coverage} (TC), but also some looser metrics that quantify either the gate values with \emph{Neuron Coverage} (NC) and \emph{Boundary Coverage} (BC), or value change in one step with \emph{Step\-wise Coverage} (SC). Details are referred to \cite{9451178}. 

Although the tool can work with any LSTM layer of a deep neural network, we consider in this tutorial the running example dataset -- Fashion-MNIST.

\section{Training or Downloading LSTMs}

The package provides some utilities to support the training of an LSTM network aimed at classifying Fashion-MNIST images.
To train a model, type the following command (the dataset itself is loaded from the @keras.datasets@ library):
% !train testrnn fashion-mnist
\begin{cmds}
python3 -m testRNN.main --model fashion_mnist --mode train
\end{cmds}
One can alternatively download a pre-trained LSTM model:
% !download testrnn fashion-mnist
\begin{cmds}
wget -P saved_models https://cgi.csc.liv.ac.uk/~acps/models/fashion_mnist_lstm.h5
\end{cmds}

\section{Template Command}

The following is the template command for \testrnn:
% !silent testrnn
\begin{cmds}
python3 -m testRNN.main --model <Model> --TestCaseNum <#Tests> --Mutation <Mutation Method> --threshold_SC <SC threshold> --threshold_BC <BC threshold> --symbols_TC <#Symbols> --seq <seq in cells to test> --mode <Mode> --outputs <OutDir>
\end{cmds}
where:
\begin{shparamdescr}
\item[<Model>] is in \{@sentiment@, @mnist@, @fashion_mnist@, @ucf101@\}. It specifies the model (and therefore dataset) to work with.

\item[<\#Tests>] specifies the expected number of test cases to be generated in a test suite.

\item[<Mutation Method>] is the test case generation algorithm, in \{@random@, @genetic@\}.
  @random@ is for random sampling method, while @genetic@ is for genetic algorithm.
  \citet{DBLP:journals/corr/abs-1911-01952} give detailed descriptions of these algorithms.
  In general, the genetic algorithm is applied once the random sampling algorithm cannot improve the coverage.
  The genetic algorithm uses the random sampling method as the basis and selects the test case according to some pre-defined fitness function.
  % Please refer to the paper for the description of the fitness function.
  In \testrnn, the fitness function is related to the coverage of test requirements.

\item[<SC threshold>] is a real number in [0, 1]. It specifies the threshold value for the stepwise coverage metric. Basically, a higher threshold represents a tighter coverage metric. Please refer to \cite{DBLP:journals/corr/abs-1911-01952} for the detailed definition of stepwise coverage metric.

\item[<BC threshold>] is a real number in [0, 1]. It specifies the threshold value for the boundary coverage metric. Basically, a higher threshold represents a tighter coverage metric. Please refer to \cite{DBLP:journals/corr/abs-1911-01952} for the detailed definition of boundary coverage metric.

\item[<\#Symbols>] is a strictly positive Integer that specifies the number of symbols to be used in the temporal coverage.  This number controls the discretization of memorised values whose temporal patterns are measured.  Basically, a greater number of symbols represents a tighter coverage metric.  Please refer to \cite{DBLP:journals/corr/abs-1911-01952} for the detailed definition of temporal coverage metric.

\item[<seq in cells to test>] depends on the models' architecture: \{@mnist@: [4, 24], @fashion_mnist@: [4, 24], @sentiment@: [400, 499], @ucf101@: [0, 10]\}

\item[<Mode>] is in \{@train@, @test@\} with default value @test@. It represents the current execution is to train an LSTM model or to test one. In the current version of the tool, the dictionary for saving and loading LSTM model files should be manually specified in the dataset class (under @TestRNN/src@).

\item[<OutDir>] specifies the path where \TestRNN saves the discovered adversarial samples, and generates the coverage report.
\end{shparamdescr}

\section{Example: Fashion-MNIST}

% \xiaowei{need more than one commands! we need to use a set of commands to setup a story. }
% NB: bootstrapping the story:
For example, we can run the following command to work with the Fashion-MNIST model with the genetic algorithm-based test case generation, and terminate when the number of test cases reaches \num{10000}.
% !testrnn fashion-mnist test
\begin{cmds}
python3 -m testRNN.main --model fashion_mnist --TestCaseNum 10000 --Mutation random --threshold_SC 0.6 --threshold_BC 0.7 --symbols_TC 3 --seq [4,24] --outputs testrnn-fm
\end{cmds}
We have manually specified the threshold parameters (@threshold_SC@, @threshold_BC@, @symbols_TC@), along with the input sequence of interest @seq@.
The above command logs its outputs into the file @testrnn-fm/record.txt@, and every generated adversarial example can be found in @testrnn-fm/adv_output@.

Contrary to \DeepConcolic, \TestRNN considers all coverage metrics at once.
The metrics notably include the standard neuron coverage (NC), which is defined as a ratio of neuron activations as in \DeepConcolic, and neuron boundary coverage (NBC) defined in terms of extreme neuron activation values.
Turing to values memorised in each LSTM cells, step-wise coverage (SC) measures the range of changes in neuron outputs over consecutive steps, whereas boundary coverage (BC) informs about extreme local behaviours of neuron outputs.
At last, temporal coverage (TC) counts all distinct temporal patterns of values memorised by the model along a specific time interval when each new input is fed into the model.
\begin{textcode}
…
128 features to be covered for NC
1280 features to be covered for KMNC
256 features to be covered for NBC
128 features to be covered for SANC
21 features to be covered for SC
21 features to be covered for BC
243 features to be covered for TC

-----------------------------------------------------
1000 samples, within which there are 149 adversarial examples
the rate of adversarial examples is 0.15

neuron coverage up to now: 1.00

KMNC up to now: 0.99

NBC up to now: 0.98

SANC up to now: 0.98

Step-wise Coverage up to now: 0.14

Boundary Coverage up to now: 0.10

Temporal Coverage up to now: 0.79

------------------------------------------------------
2000 samples, within which there are 366 adversarial examples
the rate of adversarial examples is 0.18

Test requirements are all satisfied
neuron coverage up to now: 1.00

KMNC up to now: 0.99

NBC up to now: 0.99

SANC up to now: 0.98

Step-wise Coverage up to now: 0.24

Boundary Coverage up to now: 0.33

Temporal Coverage up to now: 0.89
…
\end{textcode}
\TestRNN generates batches of \num{1000} tests and reports measures of coverage at each iteration.
We can first observe that all test requirements are already satisfied at the second iteration.
\begin{textcode}
…
------------------------------------------------------
statistics:

neuron coverage up to now: 1.00

KMNC up to now: 1.00

NBC up to now: 0.99

SANC up to now: 0.98

Step-wise Coverage up to now: 0.43

Boundary Coverage up to now: 0.71

Temporal Coverage up to now: 0.98

------------------------------------------------------
unique adv. 51
10000 samples, within which there are 1606 adversarial examples
the rate of adversarial examples is 0.16
\end{textcode}
Overall, the command above generates \num{10000} samples in less than ten minutes; in our case 51 unique adversarial examples were found.

% \nb[inline]{Some explained excerpts of \protect\inlinesh{res/logs/testrnn-outs/testrnn-fashion-mnist-test.log} would be nice right here, to explain what the tool is doing; note the run-time I got on \texttt{acps-gpu1} is ≈ 4 minutes.}

\begin{center}
  \includegraphics[width=6cm]{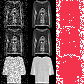}%
\end{center}

\chapter{Testing Random Forests}
\label{part:ekiml}
%(written by  Wei and Xiaowei)}

% \nb{Missing: quick introduction on random forests}%
In the \EKiML tool, we consider embedding knowledge into machine learning models.
The knowledge expression we consider can be seen as, \eg the shortcut for a targeted prediction if some features fall within the pre-defined value range, etc.
\EKiML can be used to ``\emph{embed}'' some malicious knowledge into a decision tree or random forest, representing a backdoor attack on the tree-based classifier. Particularly, black-box method, progressively adding counterexamples instead of altering of training algorithm, and the white-box method, directly modifying the saved tree structure are both supported.
The tool can also be used to ``\emph{detect}'' if a decision tree or random forest has been attacked by extracting and synthesizing the embedded knowledge. Details are referred to \cite{2020arXiv201008281H}.
% \nb[inline]{Please clearly introduce what the meaning of \emph{synthesis} is for this tool, or use \emph{detect} as suggested by Xiaowei.}
% \nb[inline]{Mention both simple trees \& forest models can be considered.}
% \nb[inline]{Also informally mention here ``\emph{black-box}'' and ``\emph{white-box}'' embedding.}
% \xiaowei{suggest change synthesis into detection, to be consistent with the paper}

\section{Training or Downloading Random Forests}

In order to exercise and demonstrate the ability of \EKiML, we provide a few pre-trained attacked models.
They can be downloaded into a @saved_models/rf-har@ sub-directory (created if needed) with the following commands:
% !download random-forests EKiML har
\begin{cmds}
wget -P saved_models/rf-har https://cgi.csc.liv.ac.uk/~acps/models/har_tree_black-box.npy
wget -P saved_models/rf-har https://cgi.csc.liv.ac.uk/~acps/models/har_forest_black-box.npy
\end{cmds}

\section{Template Command}

The following is the template command for \EKiML:
%
% !silent random-forests EKiML
\begin{cmds}
python3 -m EKiML.main --Dataset <Dataset> --Datadir <DataDir> --Mode <Mode> --Embedding_Method <EmbeddingMethod> --Model <ModelType> --Pruning <PruningFlag> --SaveModel <SaveModelFlag> --workdir <WorkDir>
\end{cmds}
where:
\begin{shparamdescr}
\item[<Dataset>] is in \{@iris@, @har@, @breast_cancer@, @mushroom@, @nursery@, @cod-rna@, @sensorless@, @mnist@\}.
  It specifies the dataset to work with, where @har@ is the dataset for our human activity recognition example;

\item[<DataDir>] is the directory where dataset files are to be found;

\item[<Mode>] is in \{@embedding@, @synthesis@\}: @synthesis@ denotes the extraction of knowledge;

\item[<EmbeddingMethod>] is in \{@black-box@, @white-box@\}. It specifies which algorithm to use for the embedding of knowledge;

\item[<ModelType>] is in \{@forest@, @tree@\}. It specifies which type of model to consider, tree or forest;

\item[<PruningFlag>] is in \{@True@, @False@\}, @False@ by default.
  It specifies if we want to apply a pre-implemented algorithm to prune the trained model.
  A pruned model can be smaller, and generalise better;

\item[<SaveModelFlag>] is in \{@True@, @False@\} @False@ by default, and specifies whether the model is saved after embedding.
  This flag is only valid in embedding mode; if @True@, the knowledge embedded tree/forest model file will be saved to the specified output directory;

\item[<WorkDir>] is the working directory, with default value
  @EKiML_workdir/@.
  In embedding mode, this specifies where the model is to be saved.  In synthesis% \red{/extraction} mode, this gives the location where
  the model is to be sought and loaded.
\end{shparamdescr}

\section{Example: Human Activity Recognition}

In the following, we provide a set of commands to work with the HAR dataset.
First, we need to manually provide the knowledge in the source file @EKiML/src/load_data.py@.
Then, one can use the following command to train a decision tree on this dataset with knowledge embedding:
% !random-forests EKiML embedding har
\begin{cmds}
python3 -m EKiML.main --Datadir datasets --Dataset har --Mode embedding --Embedding_Method black-box --Model tree --workdir saved_models/rf-har
\end{cmds}
which suggests that we are considering the HAR dataset, trying to build a decison tree by applying our black-box embedding algorithm.
Note this command may override any existing file @saved_models/rf-har/har_tree_black-box.npy@ (such as the one previously downloaded).

Example output:
\begin{textcode}
evaluation dataset:  har
embedding method:  black-box
model:  tree
trigger:  {2: -0.11, 13: 0.67, 442: -0.999}
attack label:  5
embedding Time:  44.391452805139124
No. of Training data:  7209
-----------------Pristine Classifier-------------------
No. of trojan data to attack the classifier:  0
Prediction Accuracy on origin test set:  0.9368932038834952
Prediction Accuracy on trojan test set:  0.1818770226537217
---------------Trojan Attacked Classifier---------------
No. of trojan data/paths to attack the classifier:  184
Prediction Accuracy on origin test set:  0.9284789644012945
Prediction Accuracy on trojan test set:  1.0
\end{textcode}
The output can be interpreted as follows:
We first apply the black-box method to embed the knowledge (also called trigger) $\{2: -0.11, 13: 0.67, 442: -0.999\} \Rightarrow pred = 5$ into a decision tree classifier. It costs around $44.3$ seconds and 184 counter-examples out of 7209 training data. Then the pristine and attacked classifier are evaluated in the original test set and trojan (backdoor knowledge) test set. Since the classifier achieve the $100\%$ prediction accuracy in trojan test set, the embedding is very successful with minor loss of prediction performance in orignal test set.

% !random-forests EKiML synthesis har
\begin{cmds}
python3 -m EKiML.main --Datadir datasets --Dataset har --Mode synthesis --Embedding_Method black-box --Model tree --workdir saved_models/rf-har
\end{cmds}
which suggests that we are considering the HAR dataset, trying to extract knowledge from a pre-trained tree @saved_models/rf-har/har_tree_black-box.npy@.

Example output:
\begin{textcode}
…
evaluation dataset:  har
embedding method:  black-box
Time to synthesize the knowledge:  114.96547470008954
Amount of collection:  16
Suspected features:  [array([ 52, 159, 442]), array([ 13, 429, 442]), array([ 13, 129, 442]), array([ 13,  52, 442]), array([ 52, 442]), array([ 52, 442]), array([ 52, 442]), array([ 13,  52\
, 442]), array([ 13,  52, 442]), array([ 52, 442]), array([ 52, 442]), array([ 52, 442]), array([ 52, 442]), array([ 52, 442]), array([ 13,  52, 442]), array([ 52, 159, 442])]
Suspected knowledge:  [[0.09686050587333739, 0.23038200289011002, -0.9988240003585815], [1.0, -0.9114909768104553, -0.9948055148124695], [1.0, -0.8992460072040558, -0.9896329939365387], [0.6\
774359941482544, 0.31828499608673155, -0.9989370107650757], [0.0, -0.9989795088768005], [0.0, -0.9989795088768005], [0.0, -0.9989795088768005], [0.6774359941482544, 0.31828499608673155, -0.9\
989370107650757], [0.6774359941482544, 0.31828499608673155, -0.9989370107650757], [0.0, -0.9989795088768005], [0.0, -0.9989795088768005], [0.0, -0.9989795088768005], [0.0, -0.998979508876800\
5], [0.0, -0.9989795088768005], [0.6774359941482544, 0.31828499608673155, -0.9989370107650757], [0.09686050587333739, 0.23038200289011002, -0.9988240003585815]]
\end{textcode}

From the output, it can be seen that we extract and synthesize the knowledge from the decision tree classifier. We collect 16 pieces of suspected knowledge. The knowledge includes the feature No. and their corresponding values. Based on the output, further operations, like voting can be used to summarize the most likely embedded knowledge.

\chapter{Generalizing Universal Adversarial Attacks}
%(written by Yanghao and Wenjie)}
\label{part:guap}

\GUAP implements a unified and flexible framework that can generate \emph{universal adversarial perturbation}~\cite{icdm20guap}.
Such perturbations may be applied to many inputs at the same time to produce adversarial examples.
Specifically, \GUAP can generate either additive (\ie \mbox{\(L_∞\)-bounded}) or non-additive (\eg spatial transformation) universal perturbations, or a combination of both.
This considerably generalises the attacking capability of current universal attack methods. Details are referred to \cite{icdm20guap}. 

\section{Downloading Models}

\GUAP relies on the @pytorch@ framework for constructing and
manipulating DNN models.
To exercise \GUAP, we thus provide specific pre-trained models for
Fashion-MNIST and CIFAR-10 on our server:
% !download GUAP download fashion_mnist cifar10
\begin{cmds}
wget -P saved_models https://cgi.csc.liv.ac.uk/~acps/models/fashion_mnist_modela.pth
wget -P saved_models https://cgi.csc.liv.ac.uk/~acps/models/cifar10_dense121.pth
wget -P saved_models https://cgi.csc.liv.ac.uk/~acps/models/cifar10_vgg19.pth
wget -P saved_models https://cgi.csc.liv.ac.uk/~acps/models/cifar10_resnet101.pth
\end{cmds}
\begin{table}
  \centering\smaller
  \begin{tabular}{clrr} \hline
    Layer      & Name \& Function specification             & Output shape & \#parameters    \\ \hline
    \layer 0   & \texttt{conv2d} (convolutional)            & 24 × 24 × 64 & 1664            \\
    \layer 1   & \texttt{activation} (ReLU)                 & 24 × 24 × 64 & 0               \\
    \layer 2   & \texttt{conv2d\_1} (convolutional)         & 20 × 20 × 64 & \num{102 464}   \\
    \layer 3   & \texttt{activation\_1} (ReLU)              & 20 × 20 × 64 & 0               \\
    \layer 4   & \texttt{dropout} (dropout, \(r = 0.25\))   & 20 × 20 × 64 & 0               \\
    \layer 5   & \texttt{flatten} (flat)                    & \num{25 600} & 0               \\
    \layer 6   & \texttt{dense} (dense)                     & 128          & \num{3 276 928} \\
    \layer 7   & \texttt{activation\_2} (ReLU)              & 128          & 0               \\
    \layer 8   & \texttt{dropout\_1} (dropout, \(r = 0.5\)) & 128          & 0               \\
    \layer 9   & \texttt{dense\_1} (dense)                  & 10           & \num{1 290}     \\
    \layer{10} & \texttt{activation\_3} (softmax)           & 10           & 0               \\
    \hline
  \end{tabular}
  \caption{Layers of the model
    \protect\inlinesh{fashion_mnist_modela.pth} for Fashion-MNIST}
  \label{tab:cnn-fashion-mnist-modela}
\end{table}
We show the architecture of \texttt{fashion\_mnist\_modela.pth} in \tablename~\ref{tab:cnn-fashion-mnist-modela}.  This architecture is taken from \citet{tramer2017ensemble}.

% \textcolor{red}{Yanghao: here the structure is from the paper 'Ensemble Adversarial Training: Attacks and Defenses', Table 5: \texttt{Conv-Relu-Conv-Relu-Dropout-FC-Relu-Dropout-FC-Softmax}, I can also train/transfer the same model architecture shown in Tabel~\ref{tab:cnn-fashion-mnist-medium} to make it consistent with other, let me know if needed.
% }

Some ImageNet models can be downloaded directly from the Pytorch library at \url{https://pytorch.org/docs/stable/torchvision/models.html}.

\section{Template Command}

% \nb[inline]{This is for Fashion-MNIST only; I guess the tool is not
%   generic enough, but that may do ok for now.}
% \textcolor{red}{Yanghao: I will also upload a generic version \texttt{run\_guap.py} instead, which can specify the dataset for usage.}
%
% NB: I also removed `--model modelA' as there's currently only one
% option
%
% NB: `--resume' seems broken (tries to read non-existing
% './saved_models/Fashion-MNIST_GUAP_pretrained_model.pth')
%
% !silent GUAP fashion-mnist

% \textcolor{red}{Yanghao: replace to \texttt{GUAP.run\_guap --dataset <Dataset ...>}}

\begin{cmds}
python3 -m GUAP.run_guap --dataset <Dataset> --lr <LearningRate> --batch-size <BatchSize> --epochs <Epochs> --l2reg <L2Regularization> --tau <TAU> --eps <EPSILON> --manualSeed <Seed> --gpuid <GpuList> --cuda --outdir <OutDir>
\end{cmds}
where:
\begin{shparamdescr}

\item[<Dataset>] is in \{@Fashion-MNIST@, @CIFAR10@, @IMAGENET@\}. It specifies the dataset to work with, where @Fashion-MNIST@ is our running example -- Fashion-MNIST dataset.
  % \textcolor{red}{Yanghao: add Dataset argument}

\item[<LearningRate>] is the learning rate for training the generative model (defaults to \num{0.01}).

\item[<BatchSize>] is the size for mini-batch (defaults to 100).

\item[<Epochs>] is the number of epochs to train the GUAP (default is 20).

\item[<L2Regularization>] is the weight factor for l2 regularisation (default is \num{1e-4}).

\item[<TAU>] defines the maximum magnitude for the spatial perturbation (default is \num{0.1}).

\item[<EPSILON>] defines the maximum magnitude for the \(L_∞\) noise perturbation (default is \num{0.1}).

\item[<Model>] is the name of the pre-trained model to be attacked.
  For our model for Fashion-MNIST downloaded above, this argument needs to be @modelA@. For the CIFAR10 dataset, the model needs to belong to \{@VGG19@, @ResNet101@, @DenseNet121@\}; models for ImageNet include \{@VGG16@, @VGG19@, @ResNet152@, @GoogLeNet@\}.
  % \textcolor{red}{Yanghao: add model argument}

\item[<Seed>] controls the RNG seed for the purpose of reproducibility.

\item[<GpuList>] lists the id(s) of GPU to be used, \eg @1@ or @0,1,2@% \nb{Where/how can one see the ids available on a system---a small command would be very useful}
  .

\item[--cuda] will enable CUDA for training model when this flag is present, otherwise it will use CPU only.

% \item[<Resume>] will load the pre-trained GUAP model when this
%   argument is present, otherwise GUAP will use the initial weights.

\item[<OutDir>] is the output directory, with default value @GUAP_output@, where trained \GUAP models are to be saved, as well as the % \nb{`universal flow-field and noise' would need to be introduced above I guess: representation and basic principles of computation?}
  universal flow-field and universal noise.

\item[--limited] will just use a limited amount of training data (10\%) for training model when this flag is present, otherwise it will use all training data.
  % \textcolor{red}{Yanghao: add limited argument}

\end{shparamdescr}

\section{Example: Fashion-MNIST}

We can now turn to our command for exercising the technique implemented in \GUAP on the @fashion_mnist_modela.pth@ model -- previously downloaded into @saved_models@ -- for Fashion-MNIST (note the @--limited@ flag is set to restrict the amount of training data used and obtain results in reasonable time on standard computers):
% !test GUAP fashion-mnist nocuda
\begin{cmds}
python3 -m GUAP.run_guap --dataset Fashion-MNIST --model modelA --limited
\end{cmds}
%
% \nb[inline]{An example output for the above command is given below;
%   please give some interpretation and explanations.
%   Note it took about 1h be taking too long (on CPU only—I'll see to
%   use GPUs when more resources are available) to be realistically
%   included as part of a tutorial command; may it be more realistic to
%   un-break `\mbox{\texttt{--resume}}' and provide pre-trained
%   generative models as well?
%   Or maybe using the GPU instead would be enough.}
%
% \textcolor{red}{Yanghao: For the long running time, when I use GPU (GeForce RTX 2080 Ti), it takes about 6 mins to run. I also add another argument \texttt{limited} (see above) to use a small number of for training but also can achieve comparable result. When this flag is present, it takes 40s on GPU and 3 mins on CPU (20 core). Another option is to use `\mbox{\texttt{--resume}}' and reduce the number of epochs, e.g. 5.}
%
\begin{textcode}
[2021/01/26 13:37:47] - Namespace(batch_size=100, beta1=0.5, cuda=False, dataset='Fashion-MNIST', epochs=20, eps=0.1, gpuid='0', l2reg=0.0001, lr=0.01, manualSeed=5198, model='modelA', outdir='GUAP_output', resume=False, tau=0.1)
Generalizing Universarial Adversarial Examples
==> Preparing data..
[2021/01/26 13:37:47] - Epoch 	 Time 	 Tr_loss 	 L_flow 	 Tr_acc 	 Tr_stAtt 	 Tr_noiseAtt 	 Tr_Attack Rate
[2021/01/26 13:43:36] - 0 	 348.8 	 -2.2923 	 0.1624 	 1.0000 	 0.0734 	 0.3894 	 0.6734
[2021/01/26 13:49:24] - 1 	 347.9 	 -2.5797 	 0.1471 	 1.0000 	 0.0738 	 0.4259 	 0.7390
[2021/01/26 13:55:27] - 2 	 363.6 	 -2.6761 	 0.1396 	 1.0000 	 0.0733 	 0.4499 	 0.7645
[2021/01/26 14:01:15] - 3 	 348.1 	 -2.7154 	 0.1441 	 1.0000 	 0.0730 	 0.4548 	 0.7733
[2021/01/26 14:07:01] - 4 	 345.6 	 -2.7676 	 0.1362 	 1.0000 	 0.0722 	 0.4614 	 0.7869
[2021/01/26 14:12:56] - 5 	 354.4 	 -2.7610 	 0.1364 	 1.0000 	 0.0737 	 0.4614 	 0.7844
[2021/01/26 14:18:36] - 6 	 340.1 	 -2.7781 	 0.1280 	 1.0000 	 0.0717 	 0.4657 	 0.7856
[2021/01/26 14:24:29] - 7 	 353.1 	 -2.7602 	 0.1349 	 1.0000 	 0.0721 	 0.4610 	 0.7821
[2021/01/26 14:30:24] - 8 	 355.4 	 -2.7251 	 0.1318 	 1.0000 	 0.0739 	 0.4543 	 0.7820
[2021/01/26 14:36:21] - 9 	 356.9 	 -2.7246 	 0.1349 	 1.0000 	 0.0724 	 0.4540 	 0.7781
[2021/01/26 14:42:19] - 10 	 358.4 	 -2.7317 	 0.1253 	 1.0000 	 0.0728 	 0.4632 	 0.7738
[2021/01/26 14:48:10] - 11 	 350.3 	 -2.7617 	 0.1309 	 1.0000 	 0.0732 	 0.4620 	 0.7786
[2021/01/26 14:54:09] - 12 	 359.6 	 -2.7684 	 0.1248 	 1.0000 	 0.0729 	 0.4677 	 0.7791
[2021/01/26 15:00:06] - 13 	 356.5 	 -2.7552 	 0.1210 	 1.0000 	 0.0732 	 0.4651 	 0.7789
[2021/01/26 15:06:05] - 14 	 359.6 	 -2.7615 	 0.1193 	 1.0000 	 0.0723 	 0.4658 	 0.7802
[2021/01/26 15:11:59] - 15 	 353.2 	 -2.7659 	 0.1234 	 1.0000 	 0.0732 	 0.4666 	 0.7852
[2021/01/26 15:17:53] - 16 	 354.0 	 -2.7244 	 0.1236 	 1.0000 	 0.0732 	 0.4663 	 0.7814
[2021/01/26 15:23:54] - 17 	 361.0 	 -2.7814 	 0.1252 	 1.0000 	 0.0727 	 0.4728 	 0.7896
[2021/01/26 15:29:55] - 18 	 361.5 	 -2.7367 	 0.1201 	 1.0000 	 0.0716 	 0.4634 	 0.7806
[2021/01/26 15:35:38] - 19 	 343.0 	 -2.7940 	 0.1251 	 1.0000 	 0.0720 	 0.4686 	 0.7901
Best train ASR:	0.7901166666666667
==> start testing ..
[2021/01/26 15:35:58] - Perb st Acc 	 L2 	 Time 	 Adv Test_loss 	 Te_stAtt 	 Te_noiseAtt	 Te_Attack Rate
[2021/01/26 15:35:58] - 0.2054 	 258.8305 	 19.95 	 -2.8486 	 0.0819 	 0.4937 	 0.7996
\end{textcode}
% \nb[inline]{What's relevant in the above log? What is each column
%   showing? ``ASR''?}
%
Example output image, that can be found in newly the created @GUAP_output/savefig@ directory.
\begin{center}
  \includegraphics[width=\textwidth]{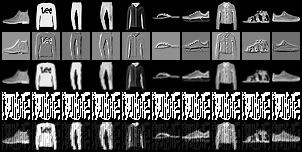}%
\end{center}
% \nb[inline]{Please explain what each row represents.}
The odd rows in the image above respectively denote:
\begin{enumerate*}[(i)]
\item original images;
\item spatial transformed images; and
\item final perturbed adversarial examples.
\end{enumerate*}
The second and fourth rows represent the differences between the original image and perturbed images caused by the spatial transform and the additive noise, respectively.
We can easily observe that the spatial-based attack mostly focuses on the areas of images where adjacent pixels show sharp changes in magnitude.
% which confirms that the edge of images plays a significant role in deep neural networks.

% ----------------------------------------------------------------------

\newpage
\bibliography{references}
\bibliographystyle{plain}

\end{document}